\documentclass[apj]{emulateapj}
\usepackage{multirow}
\usepackage{longtable}
\usepackage{rotating}

\def\gtorder{\mathrel{\raise.3ex\hbox{$>$}\mkern-14mu
             \lower0.6ex\hbox{$\sim$}}}
\def\ltorder{\mathrel{\raise.3ex\hbox{$<$}\mkern-14mu
             \lower0.6ex\hbox{$\sim$}}}

\slugcomment{Draft of \today}

\shorttitle{Asteroid rotation periods from PTF}

\shortauthors{Polishook et al.}

\begin{document}

\title{Asteroid rotation periods from the Palomar Transient Factory survey}
\author{D.~Polishook\altaffilmark{1},
E.~O.~Ofek\altaffilmark{1},
A. Waszczak\altaffilmark{2},
S.R. Kulkarni\altaffilmark{2},
A. Gal-Yam\altaffilmark{1},
O. Aharonson\altaffilmark{1},
R. Laher\altaffilmark{3},
J. Surace\altaffilmark{3},
C. Klein\altaffilmark{4},
J. Bloom\altaffilmark{4},
N. Brosch\altaffilmark{5},
D. Prialnik\altaffilmark{5},
C. Grillmair\altaffilmark{6},
S.B. Cenko\altaffilmark{4},
M. Kasliwal\altaffilmark{2},
N. Law\altaffilmark{7},
D. Levitan\altaffilmark{2},
P. Nugent\altaffilmark{8},
D. Poznanski\altaffilmark{4,5,8,9},
R. Quimby\altaffilmark{2}
}

\altaffiltext{1}{Benoziyo Center for Astrophysics, Weizmann Institute
  of Science, 76100 Rehovot, Israel.}
\altaffiltext{2}{Division of Physics, Mathematics and Astronomy,
  California Institute of Technology, Pasadena, CA 91125, USA.}
\altaffiltext{3}{Spitzer Science Center, California Institute of Technology,  M/S 314-6, Pasadena, CA 91125, U.S.A.}
\altaffiltext{4}{Department of Astronomy, University of California, Berkeley, CA, 94720-3411, USA.}
\altaffiltext{5}{Faculty of Exact Sciences, Tel-Aviv University, 69978 Tel-Aviv-Yafo, Israel.}
\altaffiltext{6}{Infrared Processing and Analysis Center, California Institute of Technology,  M/S 100-22, Pasadena, CA 91125, U.S.A.}
\altaffiltext{7}{Dunlap Institute for Astronomy and Astrophysics, University of Toronto, Toronto, M5S 3H4 Ontario, Canada.}
\altaffiltext{8}{Computational Cosmology Center, Lawrence Berkeley National Laboratory, Berkeley, CA 94720, USA.}
\altaffiltext{9}{Einstein Fellow.}

\begin{abstract}
The Palomar Transient Factory (PTF) is a synoptic survey designed to explore the transient and variable sky in a wide variety of cadences. We use PTF observations of fields that were observed multiple times ($\gtorder10$) per night, for several nights, to find asteroids, construct their lightcurves and measure their rotation periods. Here we describe the pipeline we use to achieve these goals and present the results from the first four (overlapping) PTF fields analyzed as part of this program. These fields, which cover an area of 21\,deg$^{2}$, were observed on four nights with a cadence of $\sim20$\,min. Our pipeline was able to detect 624 asteroids, of which 145 ($\approx20\%$) were previously unknown. We present high quality rotation periods for 88 main-belt asteroids and possible period or lower limit on the period for an additional 85 asteroids. For the remaining 451 asteroids, we present lower limits on their photometric amplitudes. Three of the asteroids have lightcurves that are characteristic of binary asteroids. We estimate that implementing our search for all existing high-cadence PTF data will provide rotation periods for about $10,000$ asteroids mainly in the magnitude range $\approx14$ to $\approx20$.

%        1         2         3         4         5         6         7         8   
%23456789 123456789 123456789 123456789 123456789 123456789 123456789 1234567890

\end{abstract}


\section{Introduction}
\label{sec:Introduction}
We can use time-series photometry of asteroids and other types of minor planets to study a wide variety of their physical characteristics. The rotation period (spin) can be derived from periodicity in their lightcurves (e.g., Harris et al. 1989); the lightcurve structure and changes in their mean amplitude as a function of viewing angle allow us to reconstruct their shapes (e.g., Kaasalainen \& Torppa 2001); and it provides a method to search for binary asteroids (e.g., Polishook et al. 2011). In some cases, the presence of satellites allows the determination of the mass and/or density of the asteroids (e.g., Gnat \& Sari 2010). Furthermore, and most relevant to our work, statistics of asteroid rotation periods can be used to understand the physical mechanisms that shape their rotation periods -- presumably, the two main mechanisms involved are collisions (Davis et al. 2002) and the thermal Yarkovsky�-O'Keefe-�Radzievskii-�Paddack (YORP) effect (Rubincam 2000).

%Photometry of asteroids allows the derivation of valuable data of asteroid %properties: Periodicity in lightcurves coincides with their spins; Shape is %determined by examining the lightcurve amplitude; Special features in the %lightcurves, such as eclipses, may suggest binarity and shed light on the %structure and the density of asteroids. The statistics of asteroid spin %properties enables one to study the physical mechanisms that affect the %evolution of asteroids such as collisions, planetary tidal forces and the %thermal YORP effect, in correlation with their orbits, sizes and compositions.

To date, there are $\approx3,700$ asteroids with published lightcurves and rotation periods (Warner et al. 2009). Rotation periods of asteroids are typically derived from multiple photometric observations of the same object taken on several nights. Significant contributions were made by amateur astronomers with modest equipment. In recent years, wide-field CCDs provide photometric measurements of many asteroids simultaneously, by dedicated surveys (e.g., Masiero et al. 2009; Polishook and Brosch 2009), or as part of multipurpose surveys, such as the Sloan Digital Sky Survey (SDSS; Ivezi{\' c} et al. 2001; Ofek 2011). Observations using 1--2\,m class telescopes can deliver the rotation periods for $\approx1$\,km-size main-belt asteroids. Such small asteroids are particularly interesting since the timescale of physical mechanisms, such as the YORP effect is relatively short, about $10^{6}$\,years (Rubincam 2000; Vokrouhlick{\'y} \& {\v C}apek 2002).

\begin{figure}
\centerline{\includegraphics[width=8.5cm]{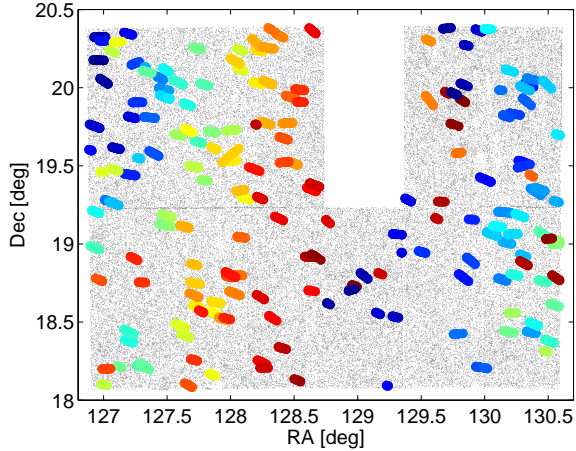}}
\caption{A single PTF field of view of 3.6$\times$2.3 deg. The colored circles are the detected asteroids while the grey dots represent stationary background sources. The missing rectangular area is due to a malfunctioning CCD. The data are based on observations of PTF field 110004 taken on Feb 14, 2010 (see Table~\ref{tab:Obs}).
\label{fig:PTFfield}}
\end{figure}

Herein we describe a pipeline designed to detect asteroids, construct their lightcurves and measure their rotation periods, in data obtained by the Palomar Transient Factory (PTF\footnote{http://www.astro.caltech.edu/ptf/}; Law et al. 2009; Rau et al. 2009). The PTF is an automated, wide-field survey aimed at a systematic exploration of the optical transient sky. Operating daily, the survey uses a camera with a 7.26\,deg$^{2}$ field of view assembled on the 1.2-m Oschin telescope at Palomar Observatory (Rahmer et al. 2008). The camera is a mosaic of 11 CCDs, with pixel scale of 1.01$''$\,pix$^{-1}$ (e.g., Figure~\ref{fig:PTFfield}). With an exposure time of 60\,s the survey reaches a limiting magnitude of $\sim21$ mag with a median seeing of $\approx2\,''$. The two filters frequently used are Mould-{\it R} and SDSS-{\it g'}. The PTF survey samples the sky in a variety of cadences to match the scientific goals of different programs, such as the search for supernovae (e.g., Arcavi et al. 2010) and galactic variables (e.g. Levitan et al. 2011). Part of the PTF time is used for high cadence observations, in which we obtain $\gtorder20$ observations of the same field in a given night, and in some cases we repeat these high cadence observations on multiple nights.

% (e.g., Arcavi et al. 2010; Ofek et al. 2010; Cooke et al. 2011; Kasliwal et al. 2011; Miller et al. 2011)

This paper is organized as follows. In \S\ref{sec:PTF-asteroids} we describe the PTF asteroid pipeline. \S\ref{sec:Observations} presents the observations analyzed in this study, while in \S\ref{sec:results} we give the results. Finally we conclude in \S\ref{sec:discussion}.

\section{The PTF asteroid pipeline}
\label{sec:PTF-asteroids}

%The asteroid identification algorithm is described in \S\ref{sec:identification}, while the photometric calibration is outlined in \S\ref{sec:photometry}. In \S\ref{sec:periods} we present the methods used to measure the asteroid rotation periods, and \S\ref{sec:caveats} discuss caveats and future improvements.

\subsection{Asteroid identification}
\label{sec:identification}

Each PTF image is processed by the IPAC-PTF\footnote{Infrared Processing and Analysis Center: http://www.ipac.caltech.edu/} pipeline. The processing includes splitting the images, de-biasing, flat-fielding, astrometric calibration, preparation of mask images, source extraction and magnitude calibration. An overview of the IPAC-PTF pipeline may be found in Grillmair et al. (2010) and is described in detail by Laher et al. (in prep.). The photometric calibration is discussed in Ofek et al. (2011a). The mask images contain information about bad pixels, probable optical ghosts, saturated pixels, aircraft/satellite tracks and more. The source detection and extraction is performed using SExtractor (Bertin and Arnouts 1996). Here we use the SExtractor MAG\_AUTO magnitudes\footnote{Defined with ${\rm kron\_fact}=1.5$ and ${\rm min\_radius}=2.5$.} (see Kron 1980). However, our experience is that the MAG\_AUTO magnitudes are slightly biased for objects fainter than 19 mag (Ofek et al. 2011a). In the future we plan to use aperture magnitude. The processed images and their associated catalog files are analyzed using the MATLAB-based PTF/Asteroids rotation pipeline described below.
% for detecting moving objects, constructing relative photometry light curves, calibrating the photometry and lightcurve analysis.

Detection of moving sources is done on a per-field and per-CCD basis in catalog space. For each PTF field/CCD, we construct a reference image based on about ten images with the best seeing values (lowest FWHM and lowest background). Then we use SExtractor to detect the sources in the reference image. We match all the sources with $2''$ matching radius in each one of the individual epochs against the reference image catalog. Matching sources against the reference catalog, rather than against one of the epochs, is done for two important reasons: (i) the moving sources do not appear in the reference catalog; and (ii) the deep reference catalogue contains almost all of the stationary sources because faint static sources which are just at, or below, the level of the background noise for a specific image will co-ad to be detected in the reference image. Therefore, even if these stationary sources are detected in only a few of the individual images they will not be mistaken as moving sources.

%and (ii) the deep reference catalog contains almost all of the stationary sources. Therefore, a faint stationary source that appears in only one image will not likely be mistaken for a moving source.

%A deep reference image is made from the field/CCD images using SWarp (Bertin et al. 2002) in order to derive only the fixed sources. A reference catalog is made out of this image and is compared to the list of sources from each image using a tolerance distance of 2\,".

The sources from each science image that do not appear in the reference catalog are considered as moving source candidates. However, a large fraction of these candidates are cosmic rays, artifacts or spurious detections. A significant part of the artifacts appear in halos around saturated stars and on  pixels created by blooming. We identify these sources by their proximity to stars brighter than 11th mag (the Tycho-2 catalog is used as a reference; H{\o}g et al. 2000) and reject them from the list of candidates.

\begin{figure}
\centerline{\includegraphics[width=8.5cm]{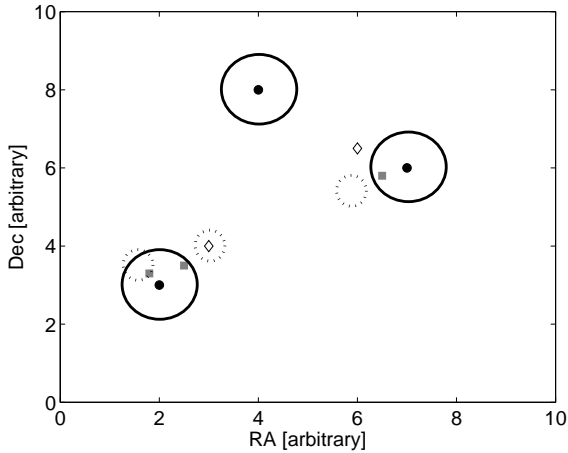}}
\caption{Illustration of the track detection algorithm. All symbols are sources that do not appear in the reference image and, therefore, they are treated as moving source candidates. The three black-filled circles appear in the first image. The program searches for candidates in the second image within a radius determined by the user and marked here by empty circles. If a candidate is found in the second image (grey squares), we search for a candidate in the third image (empty diamonds). The search region in the third image (marked by dashed circles) is calculated by Eq.~\ref{eq:searchArea}. In the case presented here only one asteroid will be detected by the code (i.e., bottom left). 
\label{fig:PTFalgorithm}}
\end{figure}

Next, we associate moving source candidates which are detections of the same physical object -- we call these "tracks". Tracks are constructed using the following algorithm: for the $i$-th candidate in the $j$-th image, the software searches for a second appearance in the $j+1$ image (the images are sorted by time). The search radius is $v_{max}(t_{j+1} - t_j)$, where $t_j$ is the time of the {\it j} image and $v_{max}$ is the maximum speed specified by the user\footnote{Main belt asteroids have typical angular speeds of $\sim0.01\,''$s$^{-1}$, while near-Earth asteroids can move at a rate of $\sim0.1\,''$s$^{-1}$ and trans-Neptunian objects have a slow motion of order 0.001\,$''$s$^{-1}$.}. Here, we use $v_{max}=0.25$\,$''$s$^{-1}$. We note that it is possible to have more than one candidate within the search radius in the $j+1$ image. Therefore, we treat each candidate source in the $j+1$ image as a possible detection of the object in the sequence. For each such possible track we look for another source in the $j+2$ image. The search is done by looking for a source around a position:
\begin{equation}
\alpha_{j+2} = \alpha_{j+1} + \frac{\alpha_{j+1} - \alpha_{j}}{t_{j+1}-t_j} (t_{j+2}-t_{j+1}) ,
\label{eq:alpha}
\end{equation}
\begin{equation}
\delta_{j+2} = \delta_{j+1} + \frac{\delta_{j+1} - \delta_{j}}{t_{j+1}-t_j} (t_{j+2}-t_{j+1}) .
\label{eq:delta}
\end{equation}
Here $\alpha$ and $\delta$ are the right ascension and the declination of the object, respectively, and the subscripts indicate the image index. The search radius, ${\it \Delta R}$, around this location is:
\begin{equation}
\Delta R = max(\Delta \alpha_{j+2}, \Delta \delta_{j+2}) + 2''
\label{eq:searchArea} ,
\end{equation}
where
\begin{equation}
\Delta \alpha_{j+2} = \sqrt{\Delta \alpha_{j+1}^2 + \frac{\Delta \alpha_{j+1}^2 + \Delta \alpha_{j}^2}{(t_{j+1}-t_j)^2} (t_{j+2}-t_{j+1})^2}
\label{eq:delta_alpha} ,
\end{equation}
\begin{equation}
\Delta \delta_{j+2} = \sqrt{\Delta \delta_{j+1}^2 + \frac{\Delta \delta_{j+1}^2 + \Delta \delta_{j}^2}{(t_{j+1}-t_j)^2} (t_{j+2}-t_{j+1})^2}
\label{eq:delta_delta} .
\end{equation}
Here $\Delta \alpha$ and $\Delta \delta$ are the astrometric errors\footnote{The astrometric errors are taken from the square of the SExtractor parameters $ERRX2WIN\_IMAGE$ and $ERRY2WIN\_IMAGE$} in the right ascension and declination, respectively. 2$''$ is a constant matching radius added in order to avoid problems due to under-estimation of the astrometric errors and any deviations from linear motion of the objects. As a result, moving sources that have a total motion of less than 2$''$ in the scanned set of images could not be detected. The search for additional sources associated with each possible track continues recursively on following images, until the moving source does not appear in the $j+n$ image. In case a single source in the first image is branching to multiple possible tracks, our recursive algorithm calculates all the possible tracks and selects the longest track (i.e., the one that contains the greatest number of points) as the most likely track. The sources that belong to this most likely track are flagged as ``associated with a track'' and the algorithm keeps running on all the sources which are not associated with a track until it runs out of sources or until the sources that remain un-associated cannot be matched. For clarity, the algorithm is visually presented in Figure~\ref{fig:PTFalgorithm}. After scanning all the images, each track is fitted with a second-order polynomial in right ascension and declination using linear least squares, and the $\chi^2$ and degrees of freedom for each fit are stored along with the track's properties.

Since the current version of the algorithm requires an asteroid to appear in successive images, it can identify multiple tracks that, in fact, belong to the same asteroid. This is possible when a moving source is not identified in one of the epochs and this may happen, for example, when the asteroid is blended with a star, or if its signal-to-noise ratio (SNR) is too low in a specific image. Therefore, next we try to merge all the tracks that belong to the same moving object. This is done by fitting the points of any pair of tracks that were observed in the same night, field and CCD to a second order polynomial. Tracks are merged if the $\chi^2$ of the fit is less than a predefined value. Similarly, single ``orphan'' sources (i.e., sources that do not appear in the reference catalog and were rejected by the detection code in previous steps) are tested as additional points belonging to the detected tracks.

Some of the tracks, especially those containing a small number of points, or those that are very short (i.e., due to small angular speed) may be artifacts and not real astrophysical sources. Another type of false alarm is created by pixels on blooming columns of saturated bright stars that were not flagged by the code in a previous stage and are mistakenly considered as sources by SExtractor. These ``objects'' have a unique trajectory (i.e., a nearly north-to-south or south-to-north motion) and have similar right ascension values to stars brighter than $\sim11$\,mag (here also we use the Tycho-2 catalog; H{\o}g et al. 2000). All the unknown detections not marked by the pipeline as false alarms are manually scanned to test their validity. While this manual scan is a simple and fast process, we wish to cancel this step by matching better thresholds to the data and make it completely automatic.

%Afterwards, the set of moving sources is scanned for false alarm detections. Typical false alarms are fixed sources with wrong astrometric solutions, and thus appear to have a proper motion. These "objects" are identified if they cross a distance that is shorter than the 2$''$ match error.

The data collected for each moving source include its position, time, angular velocity (projected into the image data), exposure parameters, $\chi^2$, number of degrees of freedom and instrumental magnitudes measured by SExtractor. Saving these values becomes handy for the remaining steps of the procedure. The detected moving sources are identified by the PyMPChecker webservice\footnote{http://dotastro.org/PyMPC/PyMPC/}. This tool provides asteroid positions around a sky and time coordinates. In addition, all the objects receive an internal PTF designation which is a running number in base 36 (digits + letters). The measured coordinates and photometry are reported to the Minor Planet Center\footnote{http://minorplanetcenter.net/}.

Asteroids observed on different nights, fields and CCDs are combined into a single track. The identification of different sources as belonging to the same asteroid is done according to the designations of the known objects. If an object is unknown, we fit a second-order polynomial to all the possible track combinations and merge tracks which are consistent with being due to a single moving source. While this method of track merging is limited to data from successive nights, it is sufficient for our purpose of spin analysis.

\subsection{Photometric calibration}
\label{sec:photometry}

In order to construct the best relative photometry lightcurve of each source we use a linear-least-squares minimization technique, in which we solve for the best zero-point normalization (per epoch), and the best ``mean'' flux density of each source that minimize the global scatter in all the lightcurves of stationary sources. Furthermore, using a set of linear constraints on the magnitudes of some of the reference stars, we simultaneously calibrate the magnitude to an absolute scale using the $r$-band magnitude of SDSS stars (York et al. 2000). This technique was introduced by Honeycutt (1992), with some modifications and the simultaneous absolute calibration presented in Ofek et al. (2011b). We note that PTF data routinely achieve relative photometry accuracy of as good as 3\,mmag (e.g., van Eyken et al. 2011; Ag\"{u}eros et al. 2011; Levitan et al. 2011).

We perform the calibration process for each night and each CCD separately. However, some asteroids are visible on multiple nights and/or multiple CCDs. Therefore, the absolute photometric accuracy of SDSS magnitudes (20\,mmag) and CCD diversity may introduce a small systematic offset between lightcurves of the same asteroid observed on multiple nights and/or multiple CCDs. We correct for these effects during the period fitting process described in \S\ref{sec:periods}.

%The photometric calibration method fits the optimal relative zero points to all of the stars of an image. This method (Ofek 2011), is performed using a linear least squares minimization that derives a photometric accuracy of 0.01 magnitudes for $\approx17$\,mag sources. Simultaneously, the data is calibrated against the Sloan Digital Sky Survey (SDSS). A second iteration is performed omitting variable stars and saturated or faint sources in order to improve the calibration accuracy. Images taken in bad weather conditions, resulting with a low number of sources, are also removed from the photometric calibration.

%While the relative calibration is resulted with a low error, the absolute calibration is less precise because the SDSS has a photometric error of about tenth of a magnitude for $\approx17$\,mag sources. This error becomes meaningful when the asteroid is measured on different chips of the CCD and on different nights because the photometric calibration is done on a chip basis. Because different reference stars have different color terms, and different sets of reference stars are used to calibrate different chips, a significant magnitude shift could interfere the combination of photometric data from different nights and chips. We decrease this magnitude shift by fitting a magnitude constant for any set of images taken on the same night/field/chip. This is done simultaneously in the rotation period search (see \S\ref{sec:periods}).

Each moving source is saved with its instrumental magnitude and zero point, for each image. Following the photometric calibration step, the observing times are corrected for light-travel time and the calibrated magnitudes ({$\it R_{PTF}$}) are reduced to geocentric ({\it r}) and heliocentric (${\it \Delta}$) distances of 1\,AU using:
\begin{equation}
M_{(r=1,\Delta=1)}= R_{PTF} + 5log(r\Delta).
\label{M}
\end{equation}

{\noindent}The mean calibrated magnitude ({$ \langle R_{PTF} \rangle $}), phase angle ($\alpha$), geocentric ({\it r}) and heliocentric (${\it \Delta}$) distances of the detected asteroids are presented in Tables~\ref{tab:ResSecure}-Table~\ref{tab:ResAmpLimit}. This correction is done only for objects with known orbital parameters.

We do not try to fit the data to a {\it H-G} system (i.e., the coefficients that describe the brightness decrease with increasing phase angle), since the phase angles of the observed asteroids change only slightly over the four-day observing period. Instead, we estimate the absolute magnitude {\it H} using a fixed {\it G} slope of 0.15 and using (Bowell et al. 1989):
\begin{equation}
H = \langle M_{(r=1,\Delta=1)} \rangle + 2.5log[(1-G){\phi}_1 + G{\phi}_2]
\label{eq_H}
\end{equation}

{\noindent}where
\begin{equation}
{\phi}_1 = exp(-3.33tan(0.5 \langle \alpha \rangle )^{0.63})
\label{eq_phi1}
\end{equation}
\begin{equation}
{\phi}_2 = exp(-1.87tan(0.5 \langle \alpha \rangle )^{1.22})
\label{eq_phi2}
\end{equation}

% and their magnitude, and/or the lightcurve's amplitude, did not change significantly during this time period.

\subsection{Rotation period analysis}
\label{sec:periods}

In order to find the synodic rotation period of the asteroids, we fit a second-order Fourier series to each lightcurve of a merged track:

% (Harris et al. 1989):
%\begin{equation}
%M=\sum_{k=1,2}{B_k\sin[\frac{2k\pi}{P}(t-t_0)]+C_k\cos[\frac{2k\pi}{P}(t-t_0)]} %+ \sum_{s=1}^{s_n}{Z_s}
%\label{eq:folding}
%\end{equation}

\begin{eqnarray}
M_j & = & \sum_{k=1,2}^{N_k}{B_k\sin[\frac{2\pi k}{P}(t_j-t_0)]+C_k\cos[\frac{2\pi k}{P}(t_j-t_0)]} \cr
  & & + \sum_{s=1}^{N_s}{Z_s} ,
\label{eq:folding}
\end{eqnarray}
where {\it $B_k$} and {\it $C_k$} are the Fourier coefficients, {\it P} is the rotation period, $M_j$ is the photometric data at $t_j$ (after the reduction to absolute planetary magnitude; i.e., Eq.~\ref{M}) and $t_0$ is an arbitrary epoch. As described in \S\ref{sec:photometry}, photometric calibration is performed separately for each set of images taken at different nights and/or CCDs. Therefore, in order to improve the photometric calibration, we also fit in Equation~\ref{eq:folding} a constant value (${\it Z_s}$) for each set of images, where a set ({\it s}) is defined as all the measurements taken on the same night, field and CCD. $N_s$ is the number of sets. For a given {\it P}, this yields a set of linear equations which is solved using least-squares minimization to obtain the free parameters. This calculation is performed for a set of trial frequencies ranging from the Nyquist frequency to one over the total time span of the lightcurve, and in steps of 0.25 divided by the time span of the lightcurve (i.e., over-sampling of 4). The typical trial periods range from about 20\,minutes to about 80\,hours, which cover the rotation periods of most asteroids (e.g., Pravec and Harris 2000).

The frequency with the minimal $\chi^2$ is chosen as the most likely period. The error in the best-fit frequency is determined by the range of periods with $\chi^2$ smaller than the minimum $\chi^2 + \Delta \chi^2$, where $\Delta \chi^2$ is calculated from the inverse $\chi^2$ distribution assuming $1 + 2N_k + N_s$ degrees of freedom. The code automatically rejects cases in which there are multiple solutions; cases where the best match is at the edge of the tested period range; or if only a few measurements exist (i.e., $<8$ data points). All the other matches are manually scanned to test the validity of their period and folded lightcurve. Lightcurves which show two peaks with photometric errors that are smaller than the lightcurve amplitude and features repeating during different nights are considered as good quality lightcurves for which a periodicity can be derived. Poor results include cases where the lightcurve is folded on the photometric noise, giving usually short periods of around 20-30\,minutes, which is the approximate sampling rate.

The folded lightcurves receive a reliability code based on the definitions of Warner et al. (2009), which are: '3' for a highly reliable result with full lightcurve coverage; '2' for an ambiguous result based on less than full coverage hence the result may be wrong by an integer ratio; and '1' for periods that are based on fragmentary or noisy lightcurves that may be completely wrong. Only periods with a reliability code of 2 or 3 are used in statistical studies of asteroid rotations. Lightcurves that contain fragmentary data with no repeating features cannot be folded. However, some of these can be used to set lower limits on the rotation periods. These limits are determined from data that show a continuous and convincing magnitude variability (the amplitude is larger than the photometric error) from a single night/CCD set that is the longest in time.

We give lower limits on the amplitude of the lightcurves of all detected asteroids. This amplitude is based on 90\% of the magnitude range, ${\Delta m}$, centered on the range median (i.e., rejecting the upper and lower 5\% of the data). This is done to avoid photometric measurements which are contaminated by nearby sources or artifacts. Since data from different night/CCD sets might have small differences in their photometric calibration, we calculate the amplitude, $A_{min}$, separately for each night/CCD set using:
\begin{equation}
A_{min} = \sqrt{{\Delta m}^2 - \langle \delta m \rangle ^2} ,
\label{MinAmp}
\end{equation}
where $\langle \delta m \rangle$ is the average photometric error. We list the median value from all the sets as the lower limit on the amplitude. We note that lightcurves with low limits on the amplitudes belong to asteroids that rotate slowly or asteroids with a nearly circular projected shape.

% (depending on the real shape and the aspect angle\footnote{The aspect angle is the angle between the line of sight and the object's spin axis.}) is near circular.

% The folded lightcurves are also manually examined for their amplitudes (defined from the minimum to the maximum of the curve). We give a lower limit on the amplitude when the coverage of the lightcurve is not complete (in case there are large gaps in phase in the folded lightcurve). In the case of lightcurves that are not folded, we give lower limits on the amplitude based on 90\% of the magnitude range (centered on the range's mode and rejecting 10\% of the data at the edges). This is done using data from a single night/chip set that has the wider range of magnitudes. The average photometric error of the relevant points is reduced from the amplitude limit (therefore cases of $Amp\geq0$ are possible). These lightcurves represent asteroids that rotate slowly or asteroids that their projectile shape (depending on the real shape and the aspect angle\footnote{The aspect angle is the angle between the line of sight and the object's spin axis.}) is mostly circular.

\subsection{Caveats and future improvements}
\label{sec:caveats}

The current algorithm misses some of the fainter asteroids. This is demonstrated in Figure~\ref{fig:MagDist} where the peak of the magnitude distribution of the detected asteroids is at brighter magnitudes than the PTF detection limit. We are able to find more moving sources, which our search algorithm failed to find, using image blinking. One of the disadvantages of the current algorithm is that, unlike MOPS-like algorithms (e.g., Grav et al. 2011), it is not searching for tracks in all possible combinations of images and that it requires the object to appear in at least three successive images.

This disadvantage is not critical for the main application presented in this paper -- rotation period measurements. This is because good photometry is available only for sources which are about one magnitude brighter than the detection limit. Such sources are rarely missed by our algorithm.

% Nevertheless, we are working on several improvements that will increase the pipeline efficiency and make it less dependent on human interaction.

%A major problem of the pipeline is the requirement to find the asteroids in at least three successive images: an asteroid which appears in five images, but crossed a fixed source on the second image and had a low {\it S/N} on the forth image is not detected by the code. These problems will be fixed in the future.

%A second issue which needs to be confronted is the further automatization of the code when it searches for the rotation period of the asteroid. So far we could not replace the experienced user by an automatic means when the true period is searched for. The automatic code can, on one hand, handle a case where the lightcurve is completely covered during a few rotations, has high {\it S/N} and a high amplitude, and on the other hand, reject cases of bad photometry with no periodicity. However, the main problem is when periodicities are found on data of mediocre quality. This is also true for calculating the amplitude, when a simple model of two harmonies can miss a deep minimum and results in a wrong amplitude value. The automatization of the pipeline will be improved with experience, and eventually we aim to run the entire algorithm without a human interference.

\section{Observations}
\label{sec:Observations}
We analyze PTF $R$-band images of four fields observed on four consecutive nights. The PTF field numbers as well as the observing dates are listed in Table~\ref{tab:Obs}. The four fields partially overlap (see Figure~\ref{fig:AllAst}) and cover a total area of 21\,deg$^{2}$. The footprints of these fields cover ecliptic latitudes between $-0.75$\,deg and $+2.5$\,deg. Table~\ref{tab:Obs} also lists the total number of exposures taken on each night and the time duration of the observing period for acquiring each set of exposures for a given night and field. The seeing in the images ranged between $2.1''$ to $2.5''$. We note that the observed fields are centered on the open cluster M44, and are also used by Ag\"{u}eros et al. (2011) to study the mass-period relation of late-K to mid-M stars.

\begin{deluxetable*}{lcccccc}
\tablecolumns{7}
\tablewidth{0pt}
\tablecaption{observed fields}
\tablehead{
%\multicolumn{1}{c}{PTF ID} &
%\multicolumn{1}{c}{RA} &
%\multicolumn{1}{c}{Dec} &
%\multicolumn{4}{c}{Number of images and total observing time} \\
%  & [deg]  & [deg] & Feb12 & Feb13 & Feb14 & Feb15
\colhead{FIELD ID} &
\colhead{RA} &
\colhead{Dec} &
\colhead{Feb12} &
\colhead{Feb13} &
\colhead{Feb14} &
\colhead{Feb15} \\
\colhead{} &
\colhead{[deg]} &
\colhead{[deg]} &
\colhead{$N_{exp}$, $\Delta t$} &
\colhead{$N_{exp}$, $\Delta t$} &
\colhead{$N_{exp}$, $\Delta t$} &
\colhead{$N_{exp}$, $\Delta t$}
}
\startdata
% PTF ID  RA           Dec     2010Feb12       2010Feb13     2010Feb14   2010Feb15
110001 & $128.75$ & $+19.25$ & $23$, 6.87 & $20$, 8.57 & $29$, 8.65 & $29$, 8.70 \\
110002 & $128.75$ & $+20.24$ & $23$, 6.87 & $19$, 8.58 & $29$, 8.65 & $29$, 8.74 \\
110003 & $131.00$ & $+19.25$ & $23$, 6.87 & $20$, 8.88 & $29$, 8.69 & $29$, 8.74 \\
110004 & $131.00$ & $+20.24$ & $23$, 6.87 & $20$, 8.88 & $29$, 8.69 & $28$, 8.78
\enddata
\tablecomments{FIELD ID is a unique number representing PTF fields. $N_{exp}$ is the number of exposures per night and field, and $\Delta t$ is the duration of time spanned by each set of observations, in hours. All observations were taken on 2010}
\label{tab:Obs}
\end{deluxetable*}

\section{Results}
\label{sec:results}

\subsection{Detected asteroids}
\label{sec:detections}

\begin{figure}
\centerline{\includegraphics[width=8.5cm]{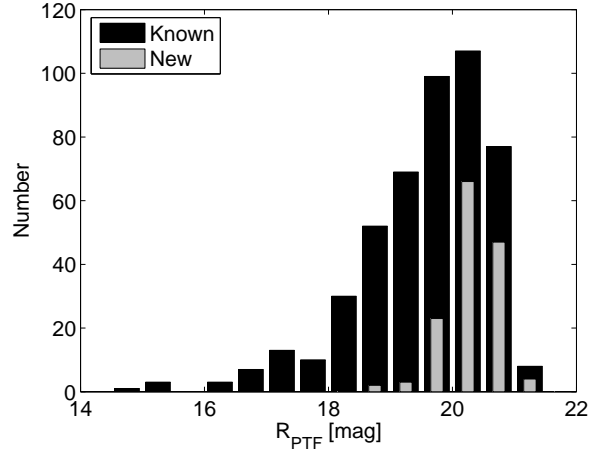}}
\caption{The distribution of magnitudes of known (black) and new (grey) asteroids found on this pilot run of the PTF survey. More than one third of the asteroids fainter than 20\,mag have not yet been discovered.
\label{fig:MagDist}}
\end{figure}

\begin{figure}
\centerline{\includegraphics[width=8.5cm]{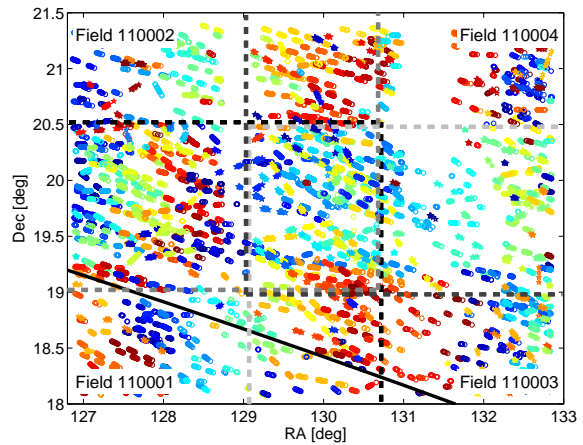}}
\caption{The tracks of 624 asteroids detected in the four overlapping fields listed in Table~\ref{tab:Obs} (marked by the dashed lines). Colored circles and pentagons are known and new asteroids, respectively. The black solid line represents the ecliptic.
\label{fig:AllAst}}
\end{figure}

We use the aforementioned asteroid pipeline to analyze the small set of PTF observations listed in Table~\ref{tab:Obs}. Running the algorithm described in \S\ref{sec:identification} we found 624 asteroids of which 145 were not yet designated by the Minor Planet Center web service. Figure~\ref{fig:AllAst} shows the tracks of all 624 detected asteroids within the field of view. Table~\ref{tab:AllMeasurements} lists all of the measurements taken for the 624 objects. The 145 unknown asteroids are fainter than 18.8 mag. The magnitude distribution, presented in Figure~\ref{fig:MagDist}, shows that more than a third of the asteroids with magnitude between 20 to 21 had not yet been discovered, while most (91\%) of the asteroids brighter than 20 are known.

\begin{figure}
\centerline{\includegraphics[width=8.5cm]{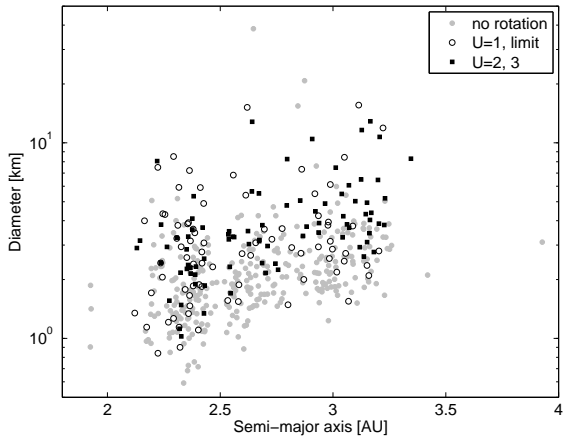}}
\caption{The estimated diameters of the known observed asteroids as a function of their semi-major axis ({\it a}). Main-belt asteroids as small as 600\,m are seen in this sample. A few bodies from the Hungaria group can be seen at $a<2\,AU$. The gap at $a=2.5\,AU$ is the 1:3 Kirkwood gap.
\label{fig:SizeDist}}
\end{figure}

The observed asteroids are located throughout the entire main-belt of asteroids, and spread out to about 4\,AU. A few members of the Hungaria group were also detected. Figure~\ref{fig:SizeDist} presents the diameters\footnote{Diameters are estimated using the formula 
\begin{equation}
D=\frac{1329}{\sqrt {\it Pv}}10^{-0.2H}
\label{D}
\end{equation}
where {\it Pv} is the geometric albedo and {\it H} is the absolute magnitude defined as the brightness of an asteroid at opposition with heliocentric and geocentric distances at 1\,AU. The estimated error of the diameters is less than a factor of two, based on a statistical error of 0.12 for {\it G} (Lagerkvist \& Magnusson 1990) and 0.1 for the albedo.} of the observed asteroids, estimated assuming a geometric albedo\footnote{the geometric albedo in visible light of main belt asteroids ranges from $\approx0.05$ to $\approx0.4$.} of 0.15, as a function of their semi-major axis ({\it a}). The plot shows that the PTF survey can detect asteroids with diameters of hundreds of meters within the inner main-belt ({\it a} $<2.5$\,AU), and 1-km size bodies in the outer main-belt.

\subsection{Derived rotation periods}
\label{sec:derived_periods}

\begin{figure*}
\centerline{\includegraphics[width=17cm]{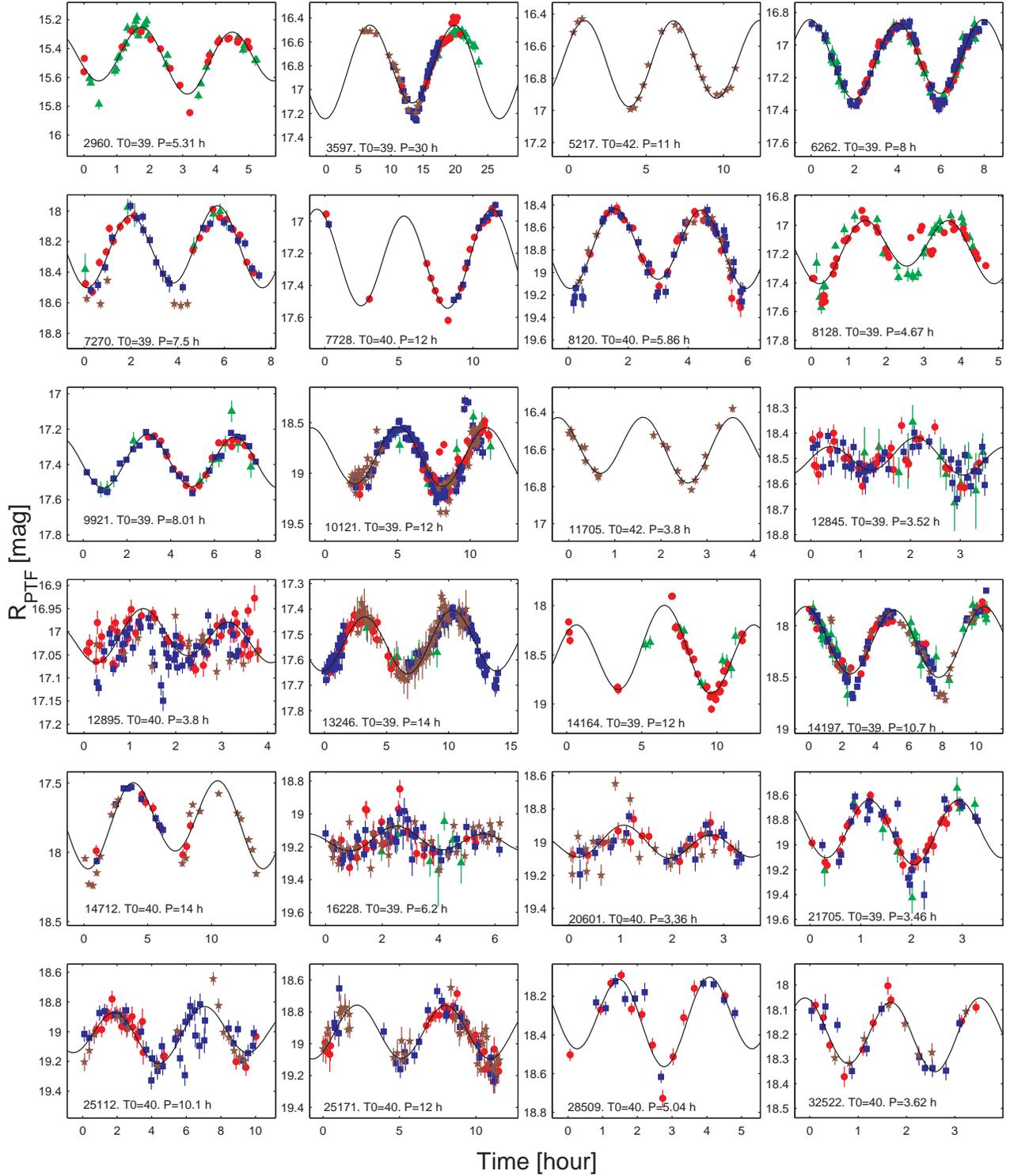}}
\caption{Set of 24 folded lightcurves. Reference epoch is $T_0+2455200$. Different markers indicate the observation date: 2010, Feb 12 - green triangles; 2010, Feb 13 - red circles; 2010, Feb 14 - blue squares; 2010, Feb 15 - brown pentagons. The designation of each asteroid, $T_0$ and the synodic rotation period {\it P} are given at the bottom of each plot.
\label{fig:LCs1}}
\end{figure*}

\begin{figure*}
\centerline{\includegraphics[width=17cm]{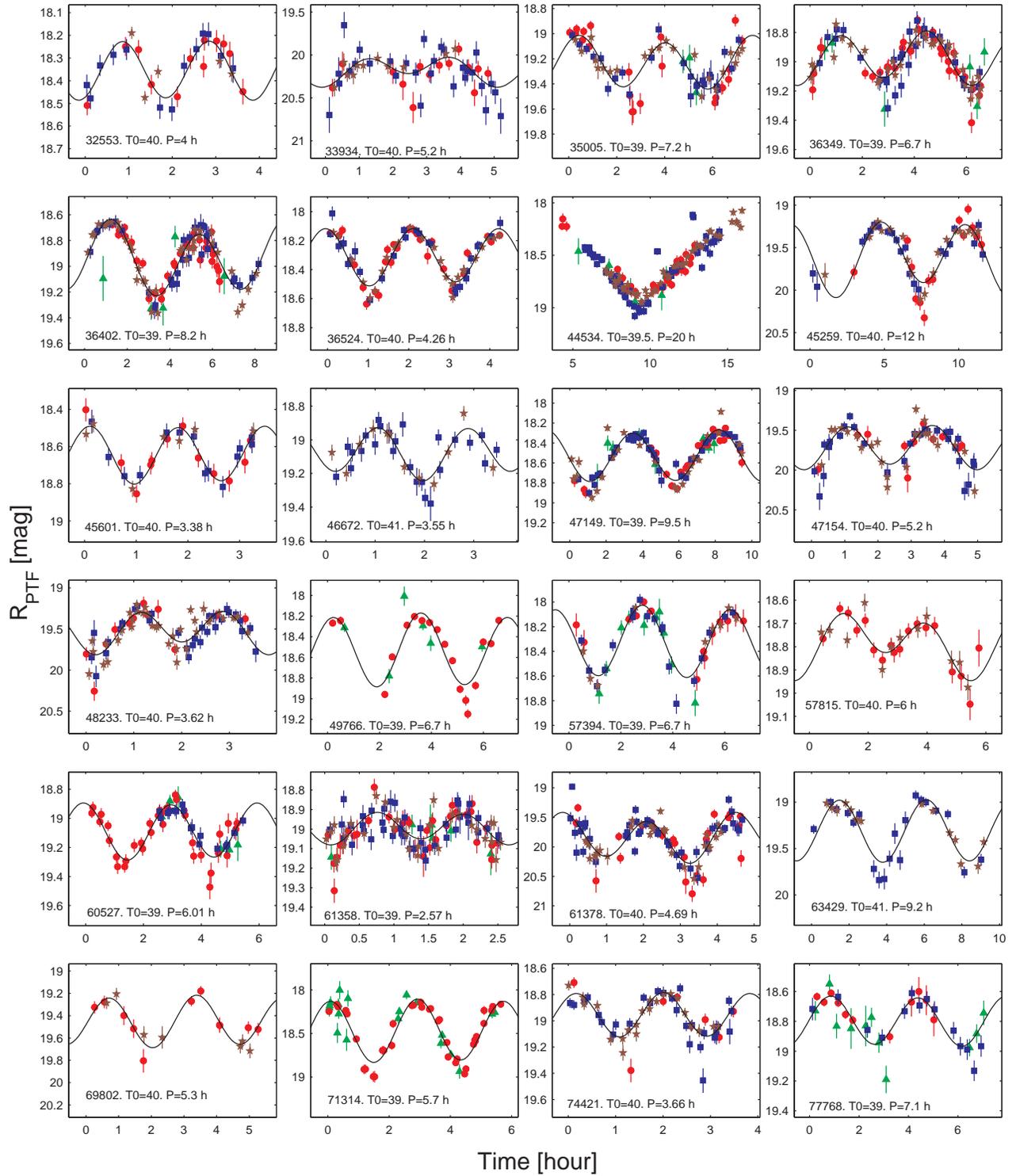}}
\caption{Same as Fig.~\ref{fig:LCs1} for 24 additional asteroids.
\label{fig:LCs2}}
\end{figure*}

\begin{figure*}
\centerline{\includegraphics[width=17cm]{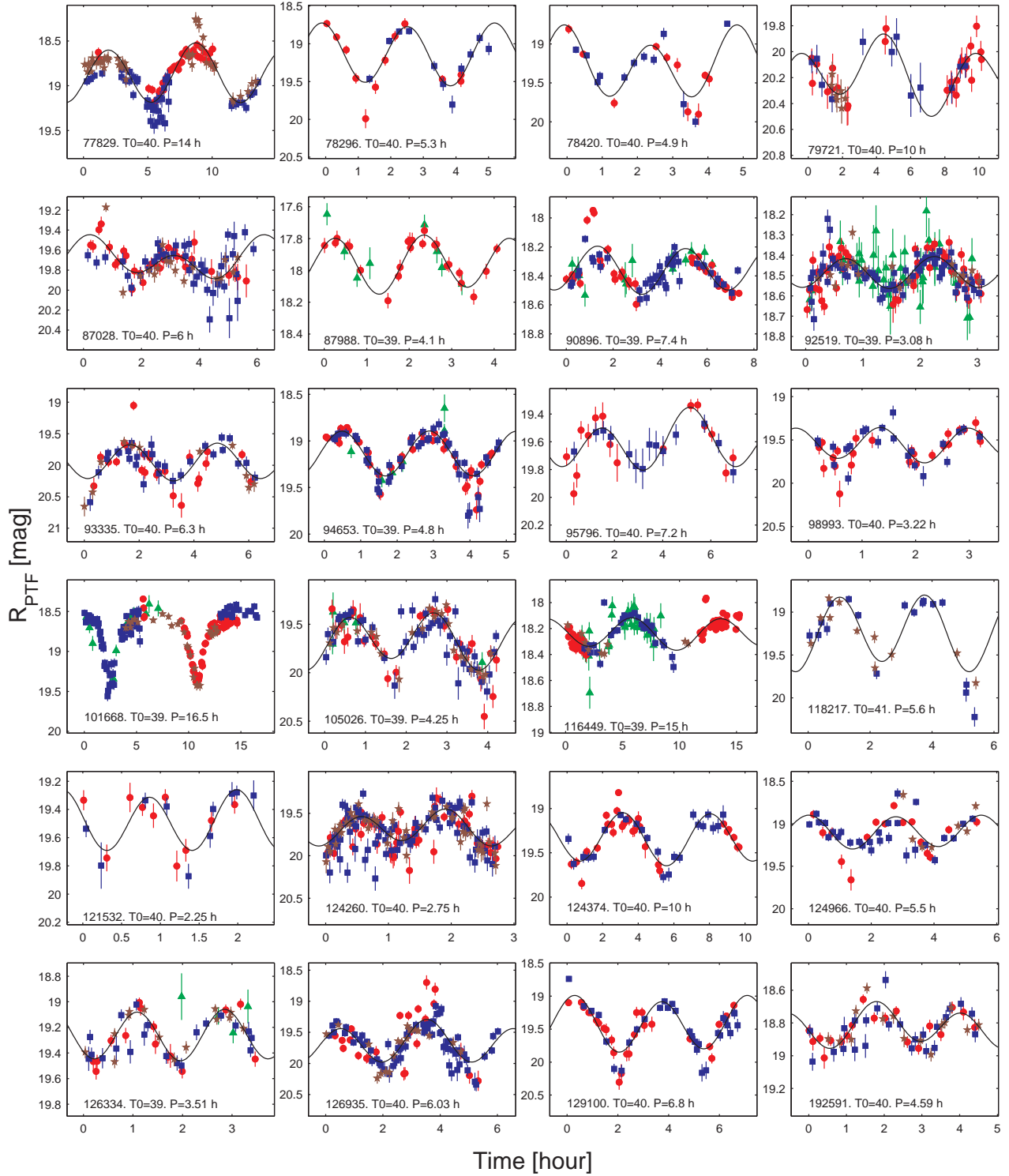}}
\caption{Same as Fig.~\ref{fig:LCs1}-\ref{fig:LCs2} for 24 additional asteroids. The deep V-shaped minima and the inverted-U-shaped maxima of asteroid (101668) 1999 {\it $CR_{95}$} suggest a binary nature.
\label{fig:LCs3}}
\end{figure*}

\begin{figure*}
\centerline{\includegraphics[width=17cm]{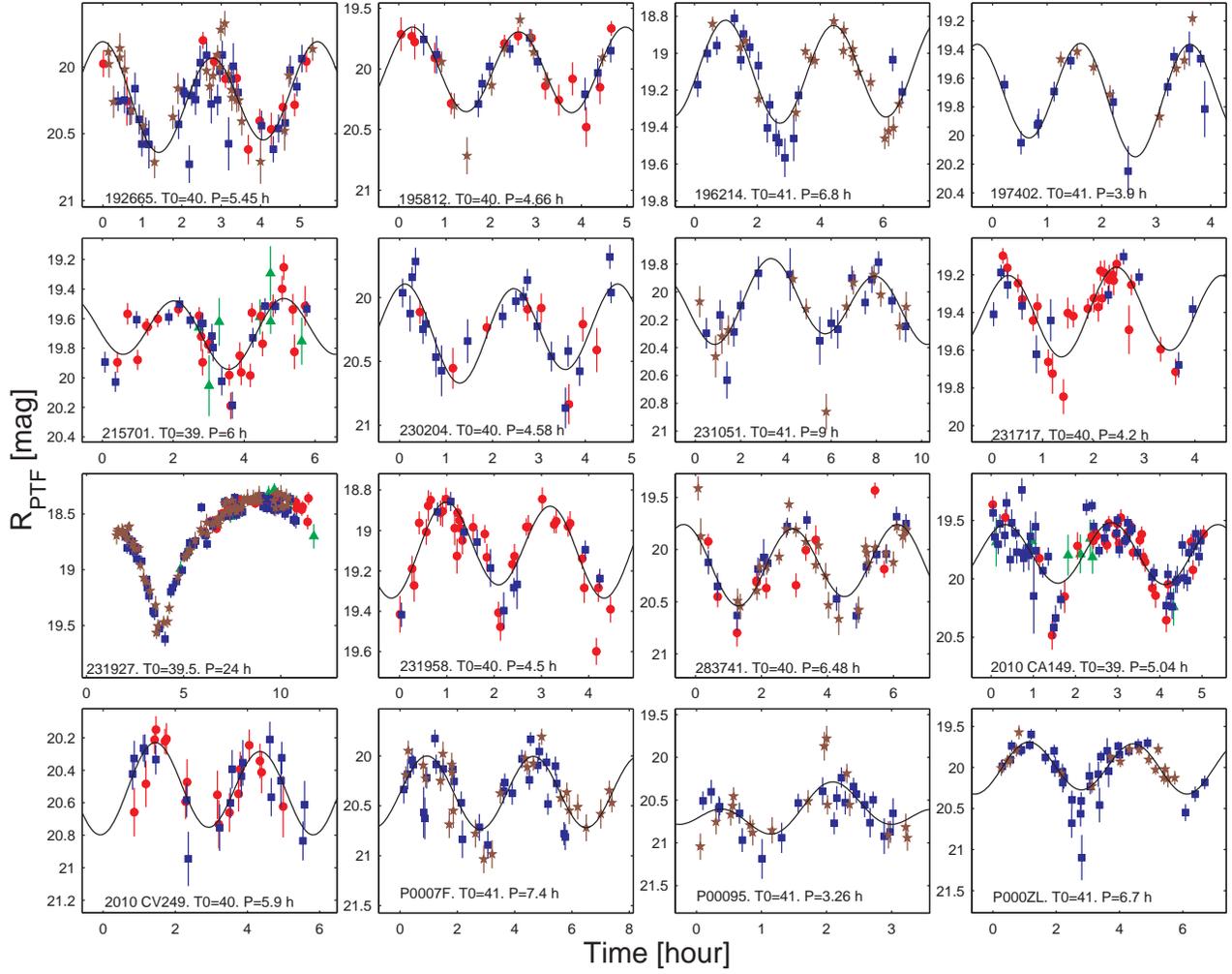}}
\caption{Same as Fig.~\ref{fig:LCs1}-\ref{fig:LCs3} for 16 additional asteroids. The deep V-shaped minima and the inverted-U-shaped maxima of asteroid (231927) 2001 {\it $DU_{30}$} suggest a binary nature.
\label{fig:LCs4}}
\end{figure*}

\begin{figure*}
\centerline{\includegraphics[width=17cm]{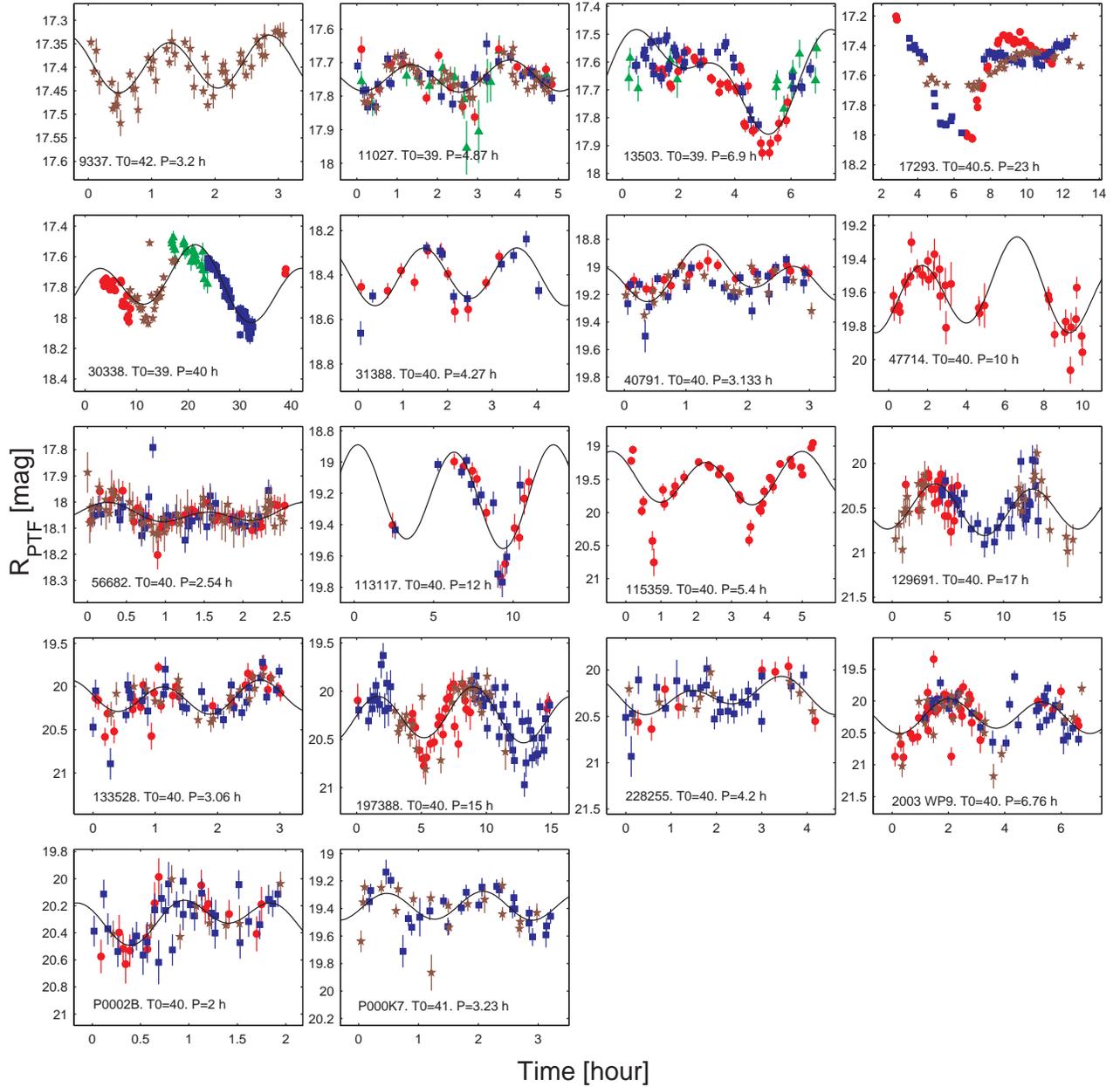}}
\caption{Set of 18 folded lightcurves with uncertain rotation periods (U=1). See Fig.~\ref{fig:LCs1} for symbols description. The deep V-shaped minima of asteroid (17293) 2743 P-L and the inconsistency of the peak's shape between different nights, suggest a binary nature.
\label{fig:LCsSuggestions1}}
\end{figure*}

\begin{figure*}
\centerline{\includegraphics[width=16cm]{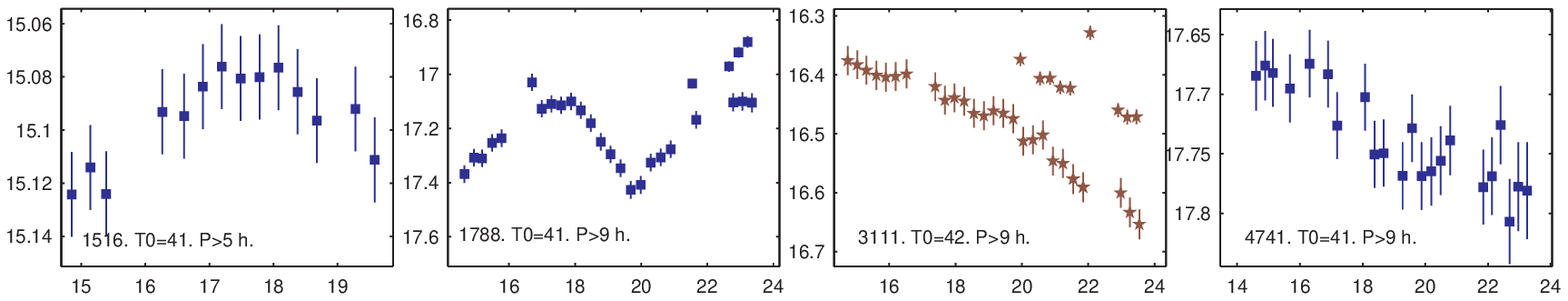}}
\caption{Set of 4 lightcurves with lower limits on the rotation periods. We only present data that show a continuous and convincing magnitude variability (the amplitude is larger than the photometric error) from a single night/CCD set that is the longest in time. See Fig.~\ref{fig:LCs1} for symbols description. This figure is published in its entirety in the electronic edition of {\it MNRAS}. A portion of the full figure is shown here for guidance regarding its form and content.
\label{fig:LCsLimits1}}
\end{figure*}

We were able to derive the rotation periods for 88 asteroids, with high reliability (codes 2 and 3), and, heretofore, none of these have a published lightcurve. The folded lightcurves are presented in Figures~\ref{fig:LCs1}$-$\ref{fig:LCs4}. The lightcurves of 85 objects with possible rotation periods and limits on the rotation period are presented in Figures~\ref{fig:LCsSuggestions1} and \ref{fig:LCsLimits1}, respectively. Four of these objects have published spin values, all are larger than the limit on the spin we present here\footnote{These four asteroids with their already published rotation periods are: 1516 Henry - 17.37\,h, 6192 1990 {\it $KB_{1}$} - 11.1\,h (both at $http://obswww.unige.ch/\sim behrend/page\_cou.html$), 10207 Comeniana - 12.84\,h (Galad et al. 2009), 23971 1998 {\it $YU_{9}$} - 6.8949\,h ($http://www.asu.cas.cz/\sim ppravec/neo.htm$).}. The rotation periods and amplitudes with high reliability for 88 asteroids are summarized in Table~\ref{tab:ResSecure}. The possible rotation periods of 18 asteroids are shown in Table~\ref{tab:ResSuggestion}, while the lower limits on the rotation periods of 67 asteroids are listed in Table~\ref{tab:ResLimit}. We also provide lower limits on the amplitude of the rest of the detected asteroids in Table~\ref{tab:ResAmpLimit}.

%These set of 118 solved lightcurves stands for almost 20\% of asteroids with rotation periods among the entire set of detected objects. 

Since most of the asteroids were observed during all four nights, where the total time span between the first and the last images taken was about 80\,hours, this gives us an opportunity to detect rotations as slow as tens of hours. Since the field of view was observed during the entire night over the course of four consecutive nights, most asteroids in our survey were observed for a total of about 32\,hours (Table~\ref{tab:Obs}). Some of the asteroids were observed for shorter times due to poor weather conditions (especially on February 12) or because they were not within the field footprints in all four nights.

% As seen in Figure~\ref{fig:MagSMA}, the PTF is sensitive enough to measure lightcurve periodicities of asteroids brighter than 20\,mag.

%\begin{figure}
%\centerline{\includegraphics[width=8.5cm]{PTF_p110001-2-3-4-_20100212-13-14-15_MagvsA.eps}}
%\caption{The calibrated $R_{PTF}$ magnitudes of the known observed asteroids as a function of their semi-major axis. The PTF is sensitive enough to measure lightcurve periodicities of asteroids brighter than 20\,mag.
%\label{fig:MagSMA}}
%\end{figure}

Among the lightcurves, three have features that are common among binary asteroids (Pravec et al. 2006). These features are deep V-shaped minima and wide inverted-U-shaped maxima. The deep V-shaped minima presumably represent a mutual event of eclipse or occultation between the primary and the secondary components. The three asteroids associated with the three aforementioned lightcurves are (17293) 2743 P-L (Figure~\ref{fig:LCsSuggestions1}); (101668) 1999 {\it $CR_{95}$} (Figure~\ref{fig:LCs3}) and (231927) 2001 {\it $DU_{30}$} (Figure~\ref{fig:LCs4}). We note that there are some inconsistencies in the shape of the lightcurve of (17293) 2743 P-L as seen on different nights. If our estimated period is valid, then this may be the result of a non-spherical shape of one of the rotating components imposed on the brightness attenuation from the mutual event. The binary nature of these three objects should be confirmed with follow-up observations. If the features in these lightcurves are indeed due to mutual events of binary asteroids, then the periods reported here are the orbital periods of the components about their center of mass, and the minima depth represents the approximate ratio between the diameters of the components.

\section{Summary and Future prospects}
\label{sec:discussion}
We present a pipeline that finds asteroids, constructs their lightcurves, and measures their rotation periods using images acquired by the PTF survey. We demonstrate our pipeline performance using a small sample of PTF data consisting of observations of about 21\,deg$^{2}$ obtained over four nights with a cadence of $\sim20$\,min. Within these images, we find 624 asteroids, of which 145 ($\cong20\%$) were previously unknown. This shows that PTF is an efficient survey for studying known, as well as unknown asteroids. Among the new discoveries, many are small, km-sized, objects. These may reveal the orbits of physically interesting objects such as small members of dynamical families (Bendjoya \& Zappal\'{a} 2002) and secondary members of rotationally-disintegrated objects (known as {\it asteroid pairs}; Vokrouhlick{\'y} \& Nesvorn{\'y} 2008).

We obtain from our analysis high quality rotation periods for 88 main-belt asteroids and additional possible periods for 85 other asteroids. Based on the ecliptic latitude distribution of PTF high-cadence observations and our success rate in measuring asteroid periodicities, we roughly estimate that PTF can measure the rotation periods for about $10,000$ asteroids. This large sample could be used to study the effects of different physical mechanisms on the spin evolution of asteroids. These mechanisms, such as collisions, the thermal YORP effect and tidal forces that follow planetary encounters, leave different signatures on the spin distribution of asteroids. For example, collisions generate a Maxwellian spin-rate distribution, as can be seen for large asteroids (\textgreater 40\,km; Pravec and Harris 2000), while the YORP effect creates a flatter spin-rate distribution (Pravec et al. 2008). A large PTF sample of asteroid rotation periods also could be used to compare the rotations of asteroids in different size groups and dynamical families (e.g., Polishook and Brosch 2009), and to track the rotational history of components ejected from rotationally-disintegrated asteroids (Pravec et al. 2010).

Dust and gas ejection from minor planets can be the result of cometary activity or a collisional event. Such objects can be detected as moving sources, with a signature extended point-spread function. The large sample of asteroids visible in PTF images is being used to search for gas and dust around minor planets and to probe their ice content, as well as to study the dynamical history of the main belt (Waszczak, Ofek \& Polishook 2011; Waszczak et al., in prep.).

PTF photometry is calibrated to a precision of better than 0.04 magnitudes, even outside SDSS footprints (Ofek et al. 2011a) and it provides accurate magnitudes of asteroids in the visible range. Therefore, PTF data could be combined with the archive of NASA's Wide-field Infrared Survey Explorer (WISE) telescope. During nine months of full cryogenic operations, the WISE telescope observed about 200,000 asteroids in the near-IR (Mainzer et al. 2010); most of these lack accurate absolute magnitudes in visible wavelengths (Masiero et al. 2011). Therefore, the two data sets combined could improve the size measurements for a significant number of asteroids.

Binary asteroids are difficult to identify because eclipses/occultations repeat themselves only after long periods (tens of hours) and last for short times. Our survey can find a large number of binary asteroids through the detection of eclipses and occultations. This can provide the statistics on binary asteroids as a function of their physical parameters. It is estimated that approximately 16\% of near-Earth asteroids are binaries (Margot et al. 2002). Applying this ratio to main-belt asteroids, and assuming that eclipses/occultations can be detected during 10\% to 30\% of the orbital period (Pravec et al. 2006), it can be estimated that out of $\approx100$ asteroids measured photometrically by the PTF, between 2 and 5 objects will register an eclipse or an occultation. This ratio is supported by our findings of three binary candidates. Enlarging the sample could determine the number of binaries among main-belt asteroids. This has important implications because any difference between the fraction of binaries among main-belt asteroids and near-Earth asteroids can constrain binary formation mechanisms and their dependencies on the asteroid environment.

% Other unique asteroids that would be detected by PTF include "tumbling" asteroids that do not rotate around a principal axis (Harris 1994), and slow and fast rotators. However, the PTF could only indicate or suggest their uniqueness and further, specifically targeted observations will be needed in order to study these objects thoroughly.

The photometry of asteroids measured by PTF could also be used to derive asteroid shapes and rotational states (sidereal period and spin axis). This is done by the lightcurve inversion method that uses brightness measurements obtained over a wide range of viewing geometries during a few apparitions to build a complete model of the asteroid (e.g., Kaasalainen and Torppa 2001, Hanu{\v s} et al. 2011). PTF will revisit many asteroids during a few apparitions, and so the lightcurve inversion method could be applied to a large sample of small-sized asteroids, thus contributing to the study of asteroid spins and shapes.

\begin{deluxetable*}{lcccccclccl}
\tablecolumns{11}
\tablewidth{0pt}
\tablecaption{Synodic rotation periods of 88 asteroids with high reliability}
\tablehead{
\colhead{Designation} &
\colhead{$N_{nights}$} &
\colhead{$N_{im}$} &
\colhead{${\it r}$} &
\colhead{$\Delta$} &
\colhead{$\alpha$} &
\colhead{Period} &
\colhead{U} &
\colhead{Amp} &
\colhead{$ \langle R_{PTF} \rangle $} &
\colhead{a} \\
\colhead{}          &
\colhead{}          &
\colhead{}          &
\colhead{[AU]}      &
\colhead{[AU]}      &
\colhead{[deg]} &
\colhead{[hour]}    &
\colhead{}          &
\colhead{[mag]}     &
\colhead{[mag]}     &
\colhead{[AU]}
}
\startdata
% Designation Nnights Nim  r Delta alpha  Rotation Period   U  Amplitude       R_{PTF}    Semi-major axis  
(2960) Ohtaki & 2 & 57 & 2 & 1.03 & 8.1 & $5.31\pm0.03$ & 3 & $\gtorder0.37$ & $15.42\pm0.02$ & 2.22 \\
(3597) Kakkuri & 4 & 97 & 3.23 & 2.28 & 5.6 & $27\pm0.5$ & 2 & $\gtorder0.58$ & $16.78\pm0.03$ & 3.17 \\
(5217) Chaozhou & 1 & 20 & 2.3 & 1.33 & 7.1 & $11.3\pm0.6$ & 2 & $\gtorder0.56$ & $16.73\pm0.01$ & 2.38 \\
(6262) Javid & 3 & 155 & 3.01 & 2.05 & 5.7 & $8.02\pm0.05$ & 3 & $\gtorder0.45$ & $17.11\pm0.03$ & 2.91 \\
(7270) Punkin & 4 & 56 & 3.79 & 2.84 & 4.8 & $7.51\pm0.07$ & 3 & $\gtorder0.5$ & $18.25\pm0.04$ & 3.21 \\
(7728) Giblin & 2 & 27 & 3.02 & 2.06 & 5.1 & $11.8\pm0.3$ & 2 & $\gtorder0.61$ & $17.18\pm0.02$ & 2.8 \\
(8120) Kobe & 3 & 79 & 2.76 & 1.81 & 6.8 & $5.86\pm0.04$ & 3 & $\gtorder0.74$ & $18.8\pm0.05$ & 2.42 \\
(8128) Nicomachus & 2 & 74 & 3.13 & 2.17 & 5.3 & $4.67\pm0.02$ & 3 & $\gtorder0.54$ & $17.19\pm0.03$ & 3.13 \\
(9921) 1981 EO18 & 3 & 55 & 2.33 & 1.36 & 6.2 & $8.01\pm0.03$ & 3 & $\gtorder0.33$ & $17.37\pm0.02$ & 2.38 \\
(10121) Arzamas & 4 & 192 & 3.67 & 2.72 & 4.7 & $12.1\pm0.3$ & 3 & $\gtorder0.66$ & $18.87\pm0.05$ & 3.2 \\
(11705) 1998 GN7 & 1 & 20 & 2.99 & 2.03 & 5.3 & $3.8\pm0.3$ & 2 & $\gtorder0.32$ & $16.65\pm0.02$ & 2.64 \\
(12845) Crick & 3 & 86 & 2.81 & 1.86 & 6.6 & $3.52\pm0.05$ & 2 & $\gtorder0.21$ & $18.51\pm0.04$ & 2.79 \\
(12895) Balbastre & 3 & 128 & 2.03 & 1.07 & 8.7 & $3.8\pm0.04$ & 2 & $\gtorder0.32$ & $17.02\pm0.02$ & 2.24 \\
(13246) 1998 MJ33 & 4 & 222 & 2.53 & 1.57 & 6.9 & $14\pm0.2$ & 3 & $\gtorder0.25$ & $17.48\pm0.03$ & 2.67 \\
(14164) Hennigar & 2 & 49 & 2.82 & 1.86 & 6.2 & $11.8\pm0.2$ & 2 & $\gtorder0.88$ & $18.54\pm0.04$ & 2.94 \\
(14197) 1998 XK72 & 4 & 161 & 3.02 & 2.06 & 5.7 & $10.7\pm0.1$ & 3 & $\gtorder0.67$ & $18.17\pm0.04$ & 3.04 \\
(14712) 2000 CO51 & 3 & 36 & 3.18 & 2.23 & 5.6 & $13.7\pm0.2$ & 2 & $\gtorder0.42$ & $17.83\pm0.03$ & 3.01 \\
(16228) 2000 EC39 & 4 & 123 & 3.6 & 2.65 & 5.3 & $6.2\pm0.2$ & 2 & $\gtorder0.27$ & $19.17\pm0.06$ & 3.12 \\
(20601) 1999 RD197 & 3 & 59 & 3.41 & 2.47 & 5.6 & $3.36\pm0.04$ & 3 & $\gtorder0.24$ & $19.02\pm0.05$ & 2.64 \\
(21705) Subinmin & 3 & 65 & 2.87 & 1.91 & 5.6 & $3.46\pm0.03$ & 3 & $\gtorder0.62$ & $18.91\pm0.06$ & 2.69 \\
(25112) Mymeshkovych & 3 & 76 & 2.94 & 1.98 & 5.7 & $10.1\pm0.2$ & 2 & $\gtorder0.44$ & $19.06\pm0.06$ & 2.88 \\
(25171) 1998 SX66 & 3 & 97 & 3.03 & 2.07 & 5.6 & $11.6\pm0.2$ & 3 & $\gtorder0.5$ & $18.99\pm0.06$ & 2.92 \\
(28509) 2000 CB92 & 2 & 23 & 2.27 & 1.3 & 6.9 & $5.04\pm0.09$ & 2 & $\gtorder0.43$ & $18.28\pm0.03$ & 2.26 \\
(32522) 2001 OE81 & 3 & 30 & 2.49 & 1.53 & 7.5 & $3.62\pm0.01$ & 3 & $\gtorder0.26$ & $18.2\pm0.03$ & 2.38 \\
(32553) 2001 QC27 & 3 & 30 & 3.53 & 2.57 & 4.3 & $4\pm0.05$ & 3 & $\gtorder0.26$ & $18.35\pm0.05$ & 3.35 \\
(33934) 2000 LA30 & 3 & 54 & 2.92 & 1.97 & 6.1 & $5.2\pm0.1$ & 2 & $\gtorder0.67$ & $20.2\pm0.1$ & 2.39 \\
(35005) 1979 MY3 & 4 & 72 & 2.83 & 1.88 & 6.6 & $7.24\pm0.06$ & 2 & $\gtorder0.47$ & $19.27\pm0.06$ & 2.56 \\
(36349) 2000 NZ23 & 4 & 104 & 2.73 & 1.77 & 6.7 & $6.7\pm0.1$ & 3 & $\gtorder0.43$ & $19.01\pm0.05$ & 2.54 \\
(36402) 2000 OT47 & 4 & 110 & 2.79 & 1.83 & 6.5 & $8.2\pm0.05$ & 3 & $\gtorder0.57$ & $18.93\pm0.05$ & 2.54 \\
(36524) 2000 QS80 & 3 & 81 & 2.5 & 1.54 & 7.1 & $4.26\pm0.03$ & 3 & $\gtorder0.44$ & $18.32\pm0.03$ & 2.56 \\
(44534) 1998 YZ9 & 4 & 144 & 2.16 & 1.21 & 9 & $22.5\pm0.5$ & 2 & $\gtorder0.7$ & $18.61\pm0.04$ & 2.37 \\
(45259) 2000 AF1 & 3 & 56 & 3.25 & 2.3 & 6.1 & $11.7\pm0.2$ & 3 & $\gtorder0.85$ & $19.54\pm0.07$ & 2.94 \\
(45601) 2000 DE5 & 3 & 33 & 3.08 & 2.12 & 4.9 & $3.38\pm0.01$ & 3 & $\gtorder0.32$ & $18.63\pm0.05$ & 2.96 \\
(46672) 1996 OA & 2 & 37 & 3.33 & 2.37 & 5.3 & $3.55\pm0.02$ & 3 & $\gtorder0.4$ & $19.08\pm0.05$ & 3.1 \\
(47149) 1999 RX34 & 4 & 100 & 2.52 & 1.56 & 7 & $9.5\pm0.09$ & 3 & $\gtorder0.49$ & $18.55\pm0.04$ & 2.13 \\
(47154) 1999 RE141 & 3 & 64 & 2.89 & 1.94 & 6.1 & $5.2\pm0.02$ & 3 & $\gtorder0.67$ & $19.71\pm0.08$ & 2.69 \\
(48233) 2001 LY9 & 3 & 79 & 3.56 & 2.61 & 5.3 & $3.62\pm0.02$ & 3 & $\gtorder0.7$ & $19.52\pm0.07$ & 3.17 \\
(49766) 1999 WS & 2 & 24 & 2.34 & 1.39 & 7.9 & $6.7\pm0.1$ & 2 & $\gtorder0.76$ & $18.5\pm0.04$ & 2.14 \\
(57394) 2001 RD84 & 4 & 52 & 2.21 & 1.24 & 7 & $6.74\pm0.06$ & 3 & $\gtorder0.6$ & $18.29\pm0.06$ & 2.32 \\
(57815) 2001 WV25 & 2 & 30 & 2.5 & 1.55 & 7.2 & $6\pm0.08$ & 2 & $\gtorder0.34$ & $18.79\pm0.04$ & 2.35 \\
(60527) 2000 EE43 & 3 & 59 & 2.97 & 2.02 & 6.5 & $6.01\pm0.04$ & 2 & $\gtorder0.36$ & $19.08\pm0.05$ & 3.05 \\
(61358) 2000 PK12 & 4 & 113 & 2.74 & 1.78 & 6.1 & $2.57\pm0.02$ & 2 & $\gtorder0.28$ & $19.01\pm0.05$ & 2.54 \\
(61378) 2000 PU28 & 3 & 103 & 2.88 & 1.92 & 6.2 & $4.69\pm0.01$ & 3 & $\gtorder0.79$ & $19.84\pm0.09$ & 2.43 \\
(63429) 2001 MH5 & 2 & 34 & 2.67 & 1.72 & 7.2 & $9.2\pm0.2$ & 3 & $\gtorder0.78$ & $19.29\pm0.06$ & 2.31 \\
(69802) 1998 RX15 & 2 & 18 & 3.01 & 2.05 & 5.8 & $5.3\pm0.1$ & 2 & $\gtorder0.48$ & $19.47\pm0.07$ & 2.87 \\
(71314) 2000 AW76 & 2 & 56 & 2.81 & 1.86 & 6.5 & $5.65\pm0.08$ & 3 & $\gtorder0.81$ & $18.45\pm0.05$ & 2.85 \\
(74421) 1999 AW24 & 3 & 58 & 2.86 & 1.91 & 6.7 & $3.66\pm0.02$ & 3 & $\gtorder0.42$ & $18.98\pm0.05$ & 3.16 \\
(77768) 2001 QM & 3 & 36 & 3.2 & 2.24 & 4.6 & $7.1\pm0.2$ & 2 & $\gtorder0.34$ & $18.77\pm0.07$ & 3.23 \\
(77829) 2001 QO217 & 3 & 125 & 3.09 & 2.14 & 5.5 & $13.6\pm0.3$ & 3 & $\gtorder0.62$ & $18.86\pm0.06$ & 3.2 \\
(78296) 2002 PT53 & 2 & 23 & 3.14 & 2.18 & 5.3 & $5.3\pm0.1$ & 3 & $\gtorder0.94$ & $19.22\pm0.07$ & 3.07 \\
(78420) 2002 QU40 & 2 & 22 & 3.44 & 2.48 & 4.5 & $4.9\pm0.05$ & 2 & $\gtorder1.11$ & $19.33\pm0.07$ & 3.07 \\
(79721) 1998 SE112 & 3 & 44 & 2.84 & 1.89 & 6.7 & $10.2\pm0.6$ & 2 & $\gtorder0.45$ & $20.2\pm0.1$ & 2.24 \\
(87028) 2000 JA78 & 3 & 93 & 2.76 & 1.81 & 6.2 & $6.02\pm0.09$ & 2 & $\gtorder0.57$ & $19.77\pm0.08$ & 2.35 \\
(87988) 2000 TZ62 & 2 & 32 & 2.25 & 1.28 & 6.6 & $4.1\pm0.2$ & 2 & $\gtorder0.42$ & $17.92\pm0.05$ & 2.62 \\
(90896) 1997 CJ3 & 3 & 91 & 2.44 & 1.48 & 7.8 & $7.4\pm0.2$ & 2 & $\gtorder0.45$ & $18.36\pm0.04$ & 2.67 \\
(92519) 2000 NO27 & 4 & 186 & 2.46 & 1.5 & 6.8 & $3.08\pm0.02$ & 3 & $\gtorder0.26$ & $18.48\pm0.05$ & 2.36 \\
(93335) 2000 SK235 & 3 & 65 & 2.42 & 1.47 & 7.5 & $6.3\pm0.1$ & 3 & $\gtorder0.9$ & $20\pm0.1$ & 2.55 \\
(94653) 2001 WF67 & 3 & 92 & 2.13 & 1.18 & 9.4 & $4.8\pm0.06$ & 3 & $\gtorder0.73$ & $19.17\pm0.06$ & 2.39 \\
(95796) 2003 FM24 & 2 & 30 & 2.54 & 1.58 & 6.1 & $7.2\pm0.2$ & 2 & $\gtorder0.45$ & $19.62\pm0.08$ & 2.39 \\
(98993) 2001 DC36 & 2 & 46 & 2.75 & 1.79 & 6.4 & $3.22\pm0.03$ & 2 & $\gtorder0.58$ & $19.6\pm0.07$ & 2.74 \\
(101668) 1999 CR95$^a$ & 4 & 178 & 2.82 & 1.87 & 6.1 & $16.54\pm0.06$ & 2 & $\gtorder0.81$ & $18.79\pm0.05$ & 3.15 \\
(105026) 2000 KX30 & 4 & 109 & 3.24 & 2.28 & 5 & $4.25\pm0.02$ & 3 & $\gtorder0.73$ & $19.68\pm0.08$ & 3.16 \\
(116449) 2004 AU & 4 & 124 & 2.61 & 1.65 & 6.5 & $15.3\pm0.7$ & 2 & $\gtorder0.33$ & $18.24\pm0.04$ & 3.15 \\
(118217) 1996 EO7 & 2 & 28 & 1.99 & 1.03 & 8.8 & $5.61\pm0.09$ & 2 & $\gtorder1.06$ & $19.27\pm0.06$ & 2.33 \\
(121532) 1999 UD40 & 2 & 20 & 2.88 & 1.92 & 5.7 & $2.25\pm0.01$ & 2 & $\gtorder0.5$ & $19.48\pm0.08$ & 2.67 \\
(124260) 2001 QK12 & 3 & 190 & 2.49 & 1.53 & 7 & $2.75\pm0.01$ & 2 & $\gtorder0.62$ & $19.73\pm0.08$ & 2.37 \\
(124374) 2001 QU152 & 2 & 51 & 2.26 & 1.3 & 8.1 & $9.7\pm0.3$ & 2 & $\gtorder0.61$ & $19.33\pm0.06$ & 2.27 \\
(124966) 2001 TH105 & 3 & 51 & 2.31 & 1.36 & 8.3 & $5.5\pm0.04$ & 2 & $\gtorder0.65$ & $19.11\pm0.05$ & 2.43 \\
(126334) 2002 AW152 & 4 & 53 & 2.42 & 1.47 & 7.7 & $3.51\pm0.02$ & 2 & $\gtorder0.43$ & $19.26\pm0.06$ & 2.54 \\
(126935) 2002 EN146 & 3 & 116 & 2.83 & 1.88 & 6.3 & $6.03\pm0.05$ & 3 & $\gtorder0.88$ & $19.7\pm0.08$ & 2.63 \\
(129100) 2004 XY4 & 2 & 57 & 2.65 & 1.69 & 6.5 & $6.8\pm0.1$ & 3 & $\gtorder0.97$ & $19.51\pm0.07$ & 2.71
\enddata
\tablecomments{Columns: asteroids' designations, number of nights, number of images, geocentric ({\it r}) and heliocentric (${\it \Delta}$) distances, phase angle ($\alpha$), rotation periods, period's reliability code, lightcurve amplitude, mean magnitude, semi-major axis.}
\tablenotetext{a}{A possible binary asteroid. The period is probably the orbital period of the satellite.}
\label{tab:ResSecure}
\end{deluxetable*}

\begin{deluxetable*}{lcccccclccl}
\setcounter{table}{2}
\tablecolumns{11}
\tablewidth{0pt}
\tablecaption{continued}
\tablehead{
\colhead{Designation} &
\colhead{$N_{nights}$} &
\colhead{$N_{im}$} &
\colhead{${\it r}$} &
\colhead{$\Delta$} &
\colhead{$\alpha$} &
\colhead{Period} &
\colhead{U} &
\colhead{Amp} &
\colhead{$ \langle R_{PTF} \rangle $} &
\colhead{a} \\
\colhead{}          &
\colhead{}          &
\colhead{}          &
\colhead{[AU]}      &
\colhead{[AU]}      &
\colhead{[deg]} &
\colhead{[hour]}    &
\colhead{}          &
\colhead{[mag]}     &
\colhead{[mag]}     &
\colhead{[AU]}
}
\startdata
% Designation Nnights Nim  r Delta alpha  Rotation Period   U  Amplitude       R_{PTF}    Semi-major axis  
(192591) 1999 CE8 & 3 & 59 & 2.59 & 1.64 & 7.8 & $4.59\pm0.03$ & 3 & $\gtorder0.36$ & $18.83\pm0.04$ & 3.15 \\
(192665) 1999 RS181 & 3 & 73 & 2.69 & 1.73 & 6.4 & $5.45\pm0.05$ & 2 & $\gtorder0.74$ & $20.2\pm0.1$ & 2.7 \\
(195812) 2002 QL19 & 3 & 36 & 3.33 & 2.37 & 4.5 & $4.66\pm0.03$ & 3 & $\gtorder0.71$ & $19.99\pm0.1$ & 3.04 \\
(196214) 2003 BD37 & 2 & 36 & 2.34 & 1.38 & 7.5 & $6.8\pm0.1$ & 2 & $\gtorder0.55$ & $19.16\pm0.06$ & 2.36 \\
(197402) 2003 YN34 & 2 & 22 & 2.71 & 1.76 & 7.1 & $3.9\pm0.02$ & 2 & $\gtorder0.93$ & $19.71\pm0.08$ & 3.14 \\
(215701) 2003 YD121 & 3 & 49 & 3.31 & 2.35 & 4.4 & $6\pm0.1$ & 2 & $\gtorder0.6$ & $19.71\pm0.08$ & 3.23 \\
(230204) 2001 ST265 & 2 & 28 & 2.16 & 1.2 & 8.4 & $4.6\pm0.1$ & 2 & $\gtorder0.66$ & $20.2\pm0.1$ & 2.43 \\
(231051) 2005 GW60 & 2 & 29 & 3.3 & 2.36 & 5.8 & $9.4\pm0.8$ & 2 & $\gtorder0.67$ & $20.2\pm0.1$ & 3.18 \\
(231717) 1999 CK9 & 2 & 36 & 2.66 & 1.7 & 6.5 & $4.2\pm0.1$ & 2 & $\gtorder0.57$ & $19.39\pm0.07$ & 3.12 \\
(231927) 2001 DU30$^a$ & 4 & 177 & 2.19 & 1.23 & 7.9 & $24.4\pm0.1$ & 2 & $\gtorder0.72$ & $18.62\pm0.05$ & 2.76 \\
(231958) 2001 PF28 & 2 & 46 & 2.89 & 1.93 & 5.6 & $4.5\pm0.1$ & 2 & $\gtorder0.6$ & $19.11\pm0.06$ & 3.06 \\
(283741) 2002 XH118 & 3 & 55 & 2.14 & 1.18 & 8.9 & $6.48\pm0.02$ & 3 & $\gtorder1.09$ & $20.1\pm0.1$ & 2.33 \\
2010 CA149 & 3 & 93 & 1.98 & 1.02 & 9.2 & $5.04\pm0.03$ & 2 & $\gtorder0.71$ & $19.76\pm0.09$ & 2.32 \\
2010 CV249 & 2 & 35 & \nodata & \nodata & \nodata & $5.9\pm0.2$ & 2 & $\gtorder0.51$ & $20.5\pm0.1$ & 3.26 \\
P0007F & 2 & 56 & \nodata & \nodata & \nodata & $7.4\pm0.3$ & 2 & $\gtorder0.93$ & $20.3\pm0.1$ & \nodata \\
P00095 & 2 & 39 & \nodata & \nodata & \nodata & $3.26\pm0.03$ & 2 & $\gtorder0.83$ & $20.6\pm0.1$ & \nodata \\
P000ZL & 2 & 53 & \nodata & \nodata & \nodata & $6.7\pm0.2$ & 2 & $\gtorder0.73$ & $20.05\pm0.1$ & \nodata
\enddata
\tablenotetext{a}{A possible binary asteroid. The period is probably the orbital period of the satellite.}
\label{tab:ResSecure2}
\end{deluxetable*}

\begin{deluxetable*}{lcccccclccl}
\tablecolumns{11}
\tablewidth{0pt}
\tablecaption{uncertain synodic rotation periods (U=1) of 18 asteroids}
\tablehead{
\colhead{Designation} &
\colhead{$N_{nights}$} &
\colhead{$N_{im}$} &
\colhead{${\it r}$} &
\colhead{$\Delta$} &
\colhead{$\alpha$} &
\colhead{Period} &
\colhead{U} &
\colhead{Amp} &
\colhead{$ \langle R_{PTF} \rangle $} &
\colhead{a} \\
\colhead{}        &
\colhead{}        &
\colhead{}        &
\colhead{[AU]}      &
\colhead{[AU]}      &
\colhead{[deg]} &
\colhead{[hour]}  &
\colhead{}        &
\colhead{[mag]}   &
\colhead{[mag]}   &
\colhead{[AU]}
}
\startdata
% Designation  Nnights Nim  r Delta alpha  Rotation Period   U  Amplitude       R_{PTF}    Semi-major axis  
(9337) 1991 FO1 & 1 & 52 & 2.8 & 1.84 & 5.6& $3.2\pm0.1$ & 1 & $\gtorder0.26$ & $17.46\pm0.02$ & 2.86 \\
(11027) Astaf'ev & 4 & 106 & 2.32 & 1.37 & 8.4& $4.87\pm0.07$ & 1 & $\gtorder0.18$ & $17.75\pm0.03$ & 2.16 \\
(13503) 1988 RH6 & 3 & 92 & 2.37 & 1.41 & 8.1& $6.9\pm0.1$ & 1 & $\gtorder0.23$ & $17.65\pm0.03$ & 2.25 \\
(17293) 2743 P-L$^a$ & 3 & 115 & 2.11 & 1.16 & 9.3& $23.24\pm0.07$ & 1 & $\gtorder0.43$ & $17.53\pm0.02$ & 2.39 \\
(30338) 2000 JW29 & 4 & 145 & 2.15 & 1.19 & 8.2& $39\pm1$ & 1 & $\gtorder0.44$ & $17.82\pm0.03$ & 2.31 \\
(31388) 1998 YL2 & 2 & 22 & 2.17 & 1.2 & 7.2& $4.27\pm0.03$ & 1 & $\gtorder0.3$ & $18.41\pm0.04$ & 2.38 \\
(40791) 1999 TO33 & 3 & 64 & 2.82 & 1.87 & 6.8& $3.133\pm0.008$ & 1 & $\gtorder0.66$ & $19.09\pm0.05$ & 2.7 \\
(47714) 2000 DS24 & 1 & 34 & 3.33 & 2.38 & 5.8& $10\pm2$ & 1 & $\gtorder0.57$ & $19.64\pm0.07$ & 2.98 \\
(56682) 2000 LA9 & 3 & 125 & 2.22 & 1.26 & 7.8& $2.54\pm0.04$ & 1 & $\gtorder0.15$ & $18.05\pm0.05$ & 2.33 \\
(113117) 2002 RF80 & 2 & 24 & 2.76 & 1.81 & 6.4& $12.3\pm0.2$ & 1 & $\gtorder0.72$ & $19.3\pm0.07$ & 3.2 \\
(115359) 2003 SA250 & 1 & 36 & 3.03 & 2.09 & 6.4& $5.4\pm0.08$ & 1 & $\gtorder1.46$ & $19.6\pm0.08$ & 2.94 \\
(129691) 1998 SH16 & 3 & 86 & 2.91 & 1.96 & 5.9& $17.1\pm0.8$ & 1 & $\gtorder0.62$ & $20.4\pm0.1$ & 2.87 \\
(133528) Ceragioli & 3 & 79 & 3.02 & 2.08 & 6.6& $3.06\pm0.04$ & 1 & $\gtorder0.52$ & $20.1\pm0.1$ & 2.86 \\
(197388) 2003 YN11 & 3 & 108 & 2.98 & 2.02 & 5.8& $14.9\pm0.9$ & 1 & $\gtorder0.83$ & $20.2\pm0.1$ & 3.15 \\
(228255) 1999 ET7 & 3 & 45 & 2.88 & 1.93 & 6.2& $4.2\pm0.1$ & 1 & $\gtorder0.54$ & $20.3\pm0.1$ & 3.16 \\
2003 WP9 & 3 & 98 & 2.9 & 1.94 & 6& $6.76\pm0.05$ & 1 & $\gtorder0.83$ & $20.2\pm0.1$ & 3.02 \\
2007 EZ184 & 3 & 54 & \nodata & \nodata & \nodata& $2\pm0.03$ & 1 & $\gtorder0.52$ & $20.3\pm0.1$ & 2.27 \\
P000K7 & 2 & 44 & \nodata & \nodata & \nodata& $3.23\pm0.02$ & 1 & $\gtorder0.47$ & $19.4\pm0.06$ & \nodata
\enddata
\tablecomments{Columns like Table~\ref{tab:ResSecure}.}
\tablenotetext{a}{A possible binary asteroid. The period is probably the orbital period of the satellite.}
\label{tab:ResSuggestion}
\end{deluxetable*}

\begin{deluxetable*}{lccccclccl}
\tablecolumns{10}
\tablewidth{0pt}
\tablecaption{Lower limits on the synodic rotation periods of 67 asteroids}
\tablehead{
\colhead{Designation} &
\colhead{$N_{nights}$} &
\colhead{$N_{im}$} &
\colhead{${\it r}$} &
\colhead{$\Delta$} &
\colhead{$\alpha$} &
\colhead{Period} &
\colhead{Amp} &
\colhead{$ \langle R_{PTF} \rangle $} &
\colhead{a} \\
\colhead{}        &
\colhead{}        &
\colhead{}        &
\colhead{[AU]}      &
\colhead{[AU]}      &
\colhead{[deg]} &
\colhead{[hour]}  &
\colhead{[mag]}   &
\colhead{[mag]}   &
\colhead{[AU]}
}
\startdata
% Designation  Nnights Nim  r Delta alpha  Rotation Period   U  Amplitude       R_{PTF}    Semi-major axis  
(1516) Henry & 2 & 23 & 2.48 & 1.51 & 6.3 & $\gtorder5$ & $\gtorder0.03$ & $15.13\pm0.02$ & 2.62 \\
(1788) Kiess & 2 & 37 & 3.61 & 2.65 & 4.9 & $\gtorder9$ & $\gtorder0.8$ & $17.29\pm0.03$ & 3.11 \\
(3111) Misuzu & 1 & 38 & 2.29 & 1.33 & 7.2 & $\gtorder9$ & $\gtorder0.18$ & $16.47\pm0.02$ & 2.22 \\
(4741) Leskov & 3 & 46 & 3.68 & 2.74 & 5.2 & $\gtorder9$ & $\gtorder0.1$ & $17.85\pm0.03$ & 3.22 \\
(5807) Mshatka & 3 & 49 & 3.37 & 2.42 & 5.4 & $\gtorder8.6$ & $\gtorder0.3$ & $18.06\pm0.04$ & 3.05
\enddata
\tablecomments{Columns: asteroids' designations, number of nights, number of images, geocentric ({\it r}) and heliocentric (${\it \Delta}$) distances, phase angle ($\alpha$), limit on the rotation period, lightcurve amplitude, mean magnitude, semi-major axis. This table is published in its entirety in the electronic edition of {\it MNRAS}. A portion of the full table is shown here for guidance regarding its form and content.}
\label{tab:ResLimit}
\end{deluxetable*}

\begin{deluxetable*}{lcccccccl}
\tablecolumns{9}
\tablewidth{0pt}
\tablecaption{Lower limits on the amplitude of 451 asteroids}
\tablehead{
\colhead{Designation} &
\colhead{$N_{nights}$} &
\colhead{$N_{im}$} &
\colhead{${\it r}$} &
\colhead{$\Delta$} &
\colhead{$\alpha$} &
\colhead{Amp} &
\colhead{$ \langle R_{PTF} \rangle $} &
\colhead{a} \\
\colhead{}        &
\colhead{}        &
\colhead{}        &
\colhead{[AU]}      &
\colhead{[AU]}      &
\colhead{[deg]} &
\colhead{[mag]}   &
\colhead{[mag]}   &
\colhead{[AU]}
}
\startdata
% Designation    Nnights Nim  r Delta alpha  Amplitude       R_{PTF}    Semi-major axis  
(625) Xenia & 3 & 57 & 3.2 & 2.25 & 5.8 & $\gtorder0$ & $14.49\pm0.02$ & 2.65 \\
(1079) Mimosa & 2 & 14 & 2.8 & 1.84 & 5.77 & $\gtorder0.1$ & $14.95\pm0.04$ & 2.87 \\
(3679) Condruses & 2 & 23 & 2.34 & 1.38 & 7.65 & $\gtorder0.1$ & $15.13\pm0.02$ & 2.2 \\
(5179) Takeshima & 2 & 37 & 2.22 & 1.26 & 7.21 & $\gtorder0.1$ & $17.29\pm0.03$ & 2.31 \\
(6916) Lewispearce & 2 & 57 & 3.2 & 2.25 & 5.34 & $\gtorder0.3$ & $15.42\pm0.02$ & 2.84
\enddata
\tablecomments{Columns: asteroids' designations, number of nights, number of images, geocentric ({\it r}) and heliocentric (${\it \Delta}$) distances, phase angle ($\alpha$), lower limit on the lightcurve amplitude, mean magnitude, semi-major axis. This table is published in its entirety in the electronic edition of {\it MNRAS}. A portion of the full table is shown here for guidance regarding its form and content.}
\label{tab:ResAmpLimit}
\end{deluxetable*}

\acknowledgments

We thank the referee for useful comments. This paper is based on observations obtained with the Samuel Oschin Telescope as part of the Palomar Transient Factory project, a scientific collaboration between the California Institute of Technology, Columbia University, Las Cumbres Observatory, the Lawrence Berkeley National Laboratory, the National Energy Research Scientific Computing Center, the University of Oxford, and the Weizmann Institute of Science. The Weizmann PTF partnership is funded in part by grants from the Israeli Science Foundation (ISF) to AG. PTF Collaborative work between the Weizmann and Caltech groups is supported by the Binational Science Foundation (BSF) via grants to SRK and AG. D Polishook further acknowledges support from the Benoziyo Center for Astrophysics and the Yeda-Sela Center at WIS. SRK and his group are partially supported by the NSF grant AST--0507734. SBC wishes to acknowledge generous support from Gary and Cynthia Bengier, the Richard and Rhoda Goldman Fund, NASA/{\it Swift} grant NNX10AI21G, NASA/{\it Fermi} grant NNX1OA057G, and National Science Foundation (NSF) grant AST--0908886.

%\begin{sidewaystable}[htdp]
\begin{sidewaystable}
\tablecolumns{16}
\tablewidth{0pt}
\centering
\caption[All measurements of the detected asteroids]{\textbf{All measurements of the detected asteroids}}
\begin{tabular}{lccccccccccccccc}
\hline \hline
Designation & Field & CCD & JD & RA & Dec & $R_{PTF}$ & AM & Background &  $ \langle FWHM  \rangle$ &  ${\it r}$ &  $\Delta$ &   $\alpha$ &  a & e & i \\
                   &         &         & [day] & [deg] & [deg] & [mag] &    &  [counts]      & [arcsec] & [AU] & [AU] & [deg] & [AU] &   & [deg] \\
\hline
(625) Xenia & 110001 & 01 & 2455239.7052872 & 127.88457 & 19.86213 & 14.4627$\pm$0.0244 & 1.15 & 1165.51 &  2.18 &  3.202 &  2.249 &   5.464 &   2.647 & 0.225 & 12.057 \\ 
(625) Xenia & 110002 & 07 & 2455239.7070572 & 127.88419 & 19.86229 & 14.5167$\pm$0.0352 & 1.14 & 1120.50 &  2.15 &  3.202 &  2.249 &   5.465 &   2.647 & 0.225 & 12.057 \\ 
(625) Xenia & 110001 & 01 & 2455239.7176472 & 127.88195 & 19.86330 & 14.4548$\pm$0.0245 & 1.12 & 1322.92 &  2.16 &  3.202 &  2.249 &   5.469 &   2.647 & 0.225 & 12.057 \\ 
(625) Xenia & 110002 & 07 & 2455239.7194372 & 127.88157 & 19.86346 & 14.5010$\pm$0.0352 & 1.11 & 1134.90 &  2.31 &  3.202 &  2.249 &   5.469 &   2.647 & 0.225 & 12.057 \\ 
(625) Xenia & 110001 & 01 & 2455239.7300672 & 127.87931 & 19.86446 & 14.4486$\pm$0.0249 & 1.09 & 2019.15 &  3.18 &  3.202 &  2.249 &   5.473 &   2.647 & 0.225 & 12.057 \\ 
\hline
\end{tabular}
\tablecomments{Columns: asteroids' designations, PTF's field and CCD, JD, RA, Dec, calibrated magnitude, air mass, image's background count and FWHM. The last six parameters are given only for objects with known orbital parameters: geocentric ({\it r}) and heliocentric (${\it \Delta}$) distances, phase angle ($\alpha$), semi-major axis, eccentricity and inclination. This table is published in its entirety in the electronic edition of {\it MNRAS}. A portion of the full table is shown here for guidance regarding its form and content.}
\label{gprop}
 \label{tab:AllMeasurements}
\end{sidewaystable}

\end{document}

\title{Asteroid rotation periods from PTF survey: Supplementary Material}

\begin{figure*}
\setcounter{figure}{10}
\centerline{\includegraphics[width=17cm]{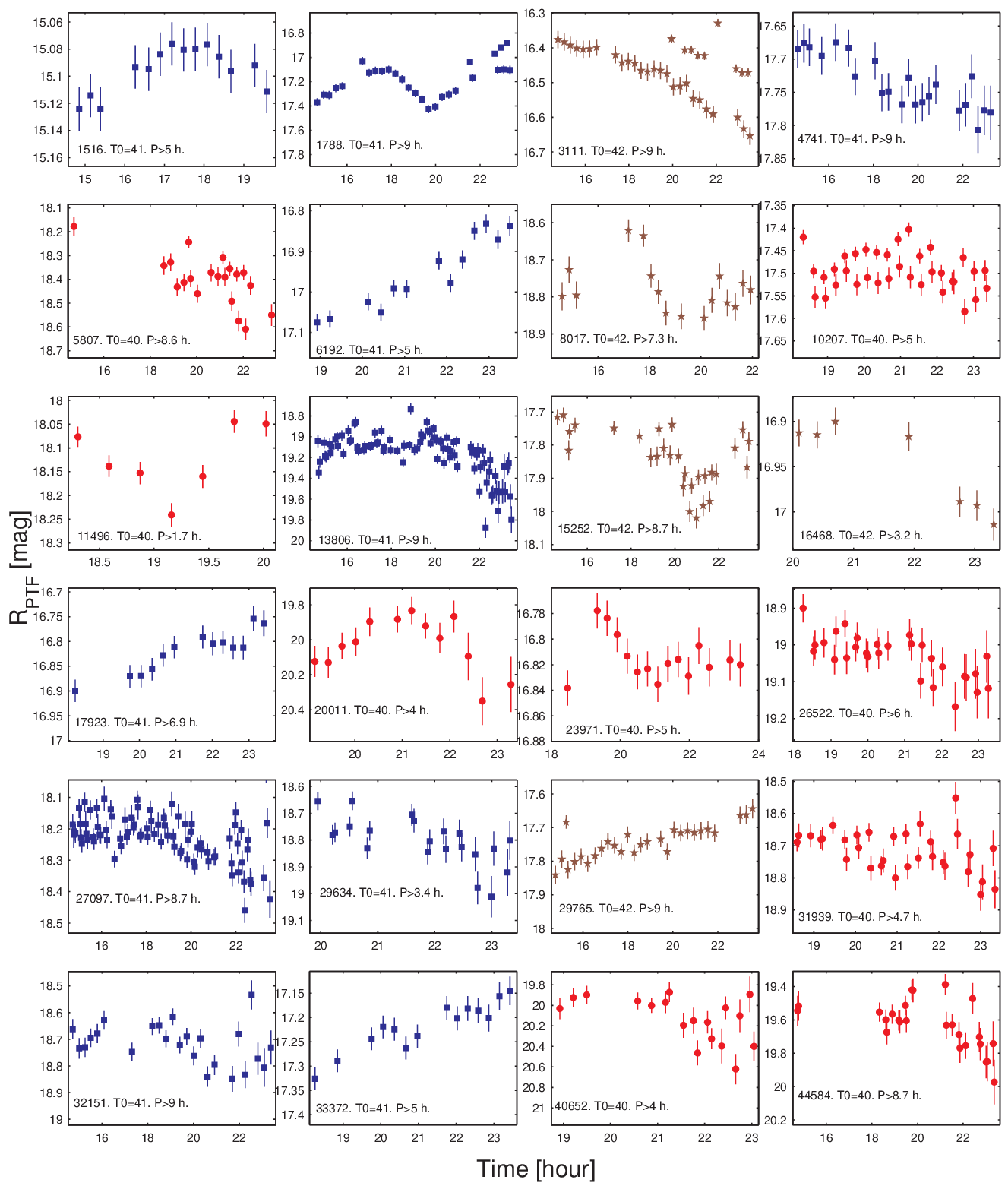}}
\caption{Set of 24 lightcurves with lower limits on the rotation periods. We only present data that show a continuous and convincing magnitude variability (the amplitude is larger than the photometric error) from a single night/CCD set that is the longest in time. Reference epoch is $T_0+2455200$. Different markers indicate the observation date: 2010, Feb 12 - green triangles; 2010, Feb 13 - red circles; 2010, Feb 14 - blue squares; 2010, Feb 15 - brown pentagons. The designation of each asteroid, $T_0$ and the synodic rotation period {\it P} are given at the bottom of each plot.
\label{fig:LCsLimits1}}
\end{figure*}

\begin{figure*}
\centerline{\includegraphics[width=17cm]{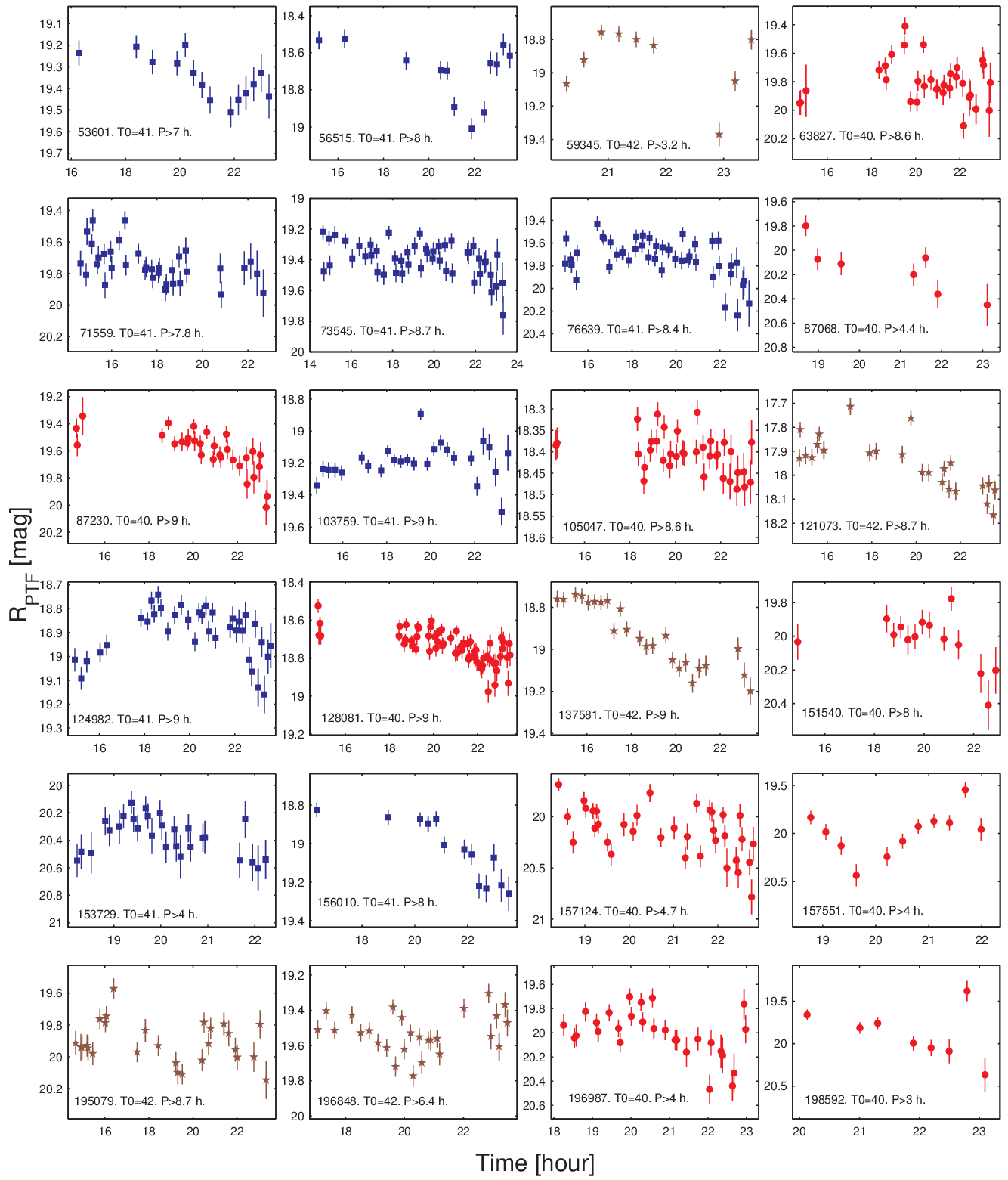}}
\caption{Same as Fig.~\ref{fig:LCsLimits1} for 24 additional asteroids. See legend in the caption of Fig.~\ref{fig:LCsLimits1}.
\label{fig:LCsLimits2}}
\end{figure*}

\begin{figure*}
\centerline{\includegraphics[width=17cm]{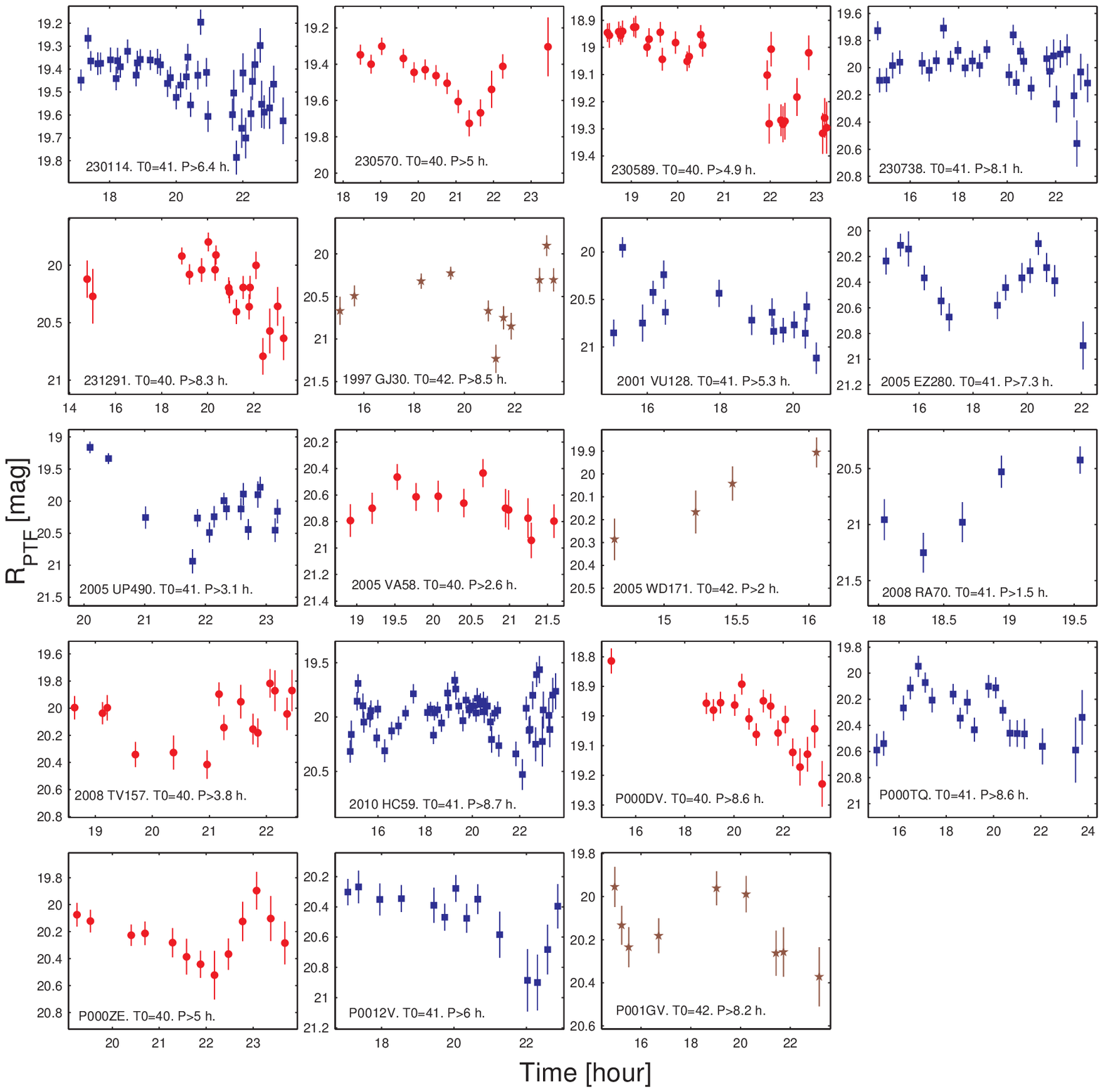}}
\caption{Same as Fig.~\ref{fig:LCsLimits1}-\ref{fig:LCsLimits2} for 19 additional asteroids. See legend in the caption of Fig.~\ref{fig:LCsLimits1}.
\label{fig:LCsLimits3}}
\end{figure*}

\begin{deluxetable*}{lccccclccl}
\setcounter{table}{4}
\tablecolumns{10}
\tablewidth{0pt}
\tablecaption{Lower limits on the synodic rotation periods of 67 asteroids}
\tablehead{
\colhead{Designation} &
\colhead{$N_{nights}$} &
\colhead{$N_{im}$} &
\colhead{${\it r}$} &
\colhead{$\Delta$} &
\colhead{$\alpha$} &
\colhead{Period} &
\colhead{Amp} &
\colhead{$ \langle R_{PTF} \rangle $} &
\colhead{a} \\
\colhead{}        &
\colhead{}        &
\colhead{}        &
\colhead{[AU]}      &
\colhead{[AU]}      &
\colhead{[deg]} &
\colhead{[hour]}  &
\colhead{[mag]}   &
\colhead{[mag]}   &
\colhead{[AU]}
}
\startdata
% Designation  Nnights Nim  r Delta alpha  Rotation Period   U  Amplitude       R_{PTF}    Semi-major axis  
(1516) Henry & 2 & 23 & 2.48 & 1.51 & 6.3 & $\gtorder5$ & $\gtorder0.03$ & $15.13\pm0.02$ & 2.62 \\
(1788) Kiess & 2 & 37 & 3.61 & 2.65 & 4.9 & $\gtorder9$ & $\gtorder0.8$ & $17.29\pm0.03$ & 3.11 \\
(3111) Misuzu & 1 & 38 & 2.29 & 1.33 & 7.2 & $\gtorder9$ & $\gtorder0.18$ & $16.47\pm0.02$ & 2.22 \\
(4741) Leskov & 3 & 46 & 3.68 & 2.74 & 5.2 & $\gtorder9$ & $\gtorder0.1$ & $17.85\pm0.03$ & 3.22 \\
(5807) Mshatka & 3 & 49 & 3.37 & 2.42 & 5.4 & $\gtorder8.6$ & $\gtorder0.3$ & $18.06\pm0.04$ & 3.05 \\
(6192) 1990 KB1 & 2 & 25 & 2.82 & 1.86 & 5.8 & $\gtorder5$ & $\gtorder0.15$ & $16.85\pm0.03$ & 2.29 \\
(8017) 1990 RM5 & 4 & 106 & 2.87 & 1.92 & 6.9 & $\gtorder7.3$ & $\gtorder0.29$ & $18.84\pm0.05$ & 2.36 \\
(10207) Comeniana & 3 & 97 & 2.58 & 1.63 & 7.5 & $\gtorder5$ & $\gtorder0.1$ & $17.51\pm0.03$ & 2.41 \\
\enddata
\tablecomments{Columns: asteroids' designations, number of nights, number of images, geocentric ({\it r}) and heliocentric (${\it \Delta}$) distances, phase angle ($\alpha$), limit on the rotation period, lightcurve amplitude, mean magnitude, semi-major axis.}
\label{tab:ResLimit}
\end{deluxetable*}

\begin{deluxetable*}{lcccccccl}
\setcounter{table}{5}
\tablecolumns{9}
\tablewidth{0pt}
\tablecaption{Lower limits on the amplitude of 451 asteroids}
\tablehead{
\colhead{Designation} &
\colhead{$N_{nights}$} &
\colhead{$N_{im}$} &
\colhead{${\it r}$} &
\colhead{$\Delta$} &
\colhead{$\alpha$} &
\colhead{Amp} &
\colhead{$ \langle R_{PTF} \rangle $} &
\colhead{a} \\
\colhead{}        &
\colhead{}        &
\colhead{}        &
\colhead{[AU]}      &
\colhead{[AU]}      &
\colhead{[deg]} &
\colhead{[mag]}   &
\colhead{[mag]}   &
\colhead{[AU]}
}
\startdata
% Designation    Nnights Nim  r Delta alpha  Amplitude       R_{PTF}    Semi-major axis  
(625) Xenia & 3 & 57 & 3.2 & 2.25 & 5.8 & $\gtorder0$ & $14.49\pm0.02$ & 2.65 \\
(1079) Mimosa & 2 & 14 & 2.8 & 1.84 & 5.77 & $\gtorder0.1$ & $14.95\pm0.04$ & 2.87 \\
(3679) Condruses & 2 & 23 & 2.34 & 1.38 & 7.65 & $\gtorder0.1$ & $15.13\pm0.02$ & 2.2 \\
(5179) Takeshima & 2 & 37 & 2.22 & 1.26 & 7.21 & $\gtorder0.1$ & $17.29\pm0.03$ & 2.31 \\
(6916) Lewispearce & 2 & 57 & 3.2 & 2.25 & 5.34 & $\gtorder0.3$ & $15.42\pm0.02$ & 2.84 \\
(12992) 1981 EZ22 & 1 & 38 & 2.61 & 1.66 & 7.03 & $\gtorder0.3$ & $16.47\pm0.02$ & 2.21 \\
(13375) 1998 VH26 & 4 & 97 & 2.88 & 1.92 & 6.01 & $\gtorder0.1$ & $16.78\pm0.03$ & 2.93 \\
(13743) Rivkin & 1 & 10 & 2.56 & 1.6 & 6.96 & $\gtorder0.2$ & $17.2\pm0.03$ & 2.24 \\
\enddata
\tablecomments{Columns: asteroids' designations, number of nights, number of images, geocentric ({\it r}) and heliocentric (${\it \Delta}$) distances, phase angle ($\alpha$), lower limit on the lightcurve amplitude, mean magnitude, semi-major axis.}
\label{tab:ResAmpLimit}
\end{deluxetable*}

\begin{deluxetable*}{lcccccccl}
\setcounter{table}{5}
\tablecolumns{9}
\tablewidth{0pt}
\tablecaption{continued}
\tablehead{
\colhead{Designation} &
\colhead{$N_{nights}$} &
\colhead{$N_{im}$} &
\colhead{${\it r}$} &
\colhead{$\Delta$} &
\colhead{$\alpha$} &
\colhead{Amp} &
\colhead{$ \langle R_{PTF} \rangle $} &
\colhead{a} \\
\colhead{}        &
\colhead{}        &
\colhead{}        &
\colhead{[AU]}      &
\colhead{[AU]}      &
\colhead{[deg]} &
\colhead{[mag]}   &
\colhead{[mag]}   &
\colhead{[AU]}
}
\startdata
% Designation    Nnights Nim  r Delta alpha  Amplitude       R_{PTF}    Semi-major axis  
(91455) 1999 RV49 & 3 & 56 & 2.41 & 1.44 & 6.49 & $\gtorder0.3$ & $18.71\pm0.04$ & 2.67 \\
(94798) 2001 XL155 & 3 & 50 & 2.72 & 1.76 & 6.39 & $\gtorder0.5$ & $17.25\pm0.03$ & 2.43 \\
(96532) 1998 RH69 & 3 & 54 & 2.48 & 1.53 & 7.96 & $\gtorder0.3$ & $20.2\pm0.1$ & 2.98 \\
(97681) 2000 GA9 & 4 & 72 & 2.55 & 1.59 & 7.41 & $\gtorder0.3$ & $19.27\pm0.06$ & 3 \\
(97777) 2000 KG16 & 4 & 104 & 3.7 & 2.74 & 4.2 & $\gtorder0.3$ & $19.01\pm0.05$ & 3.22 \\
(99698) 2002 JY31 & 4 & 110 & 2.47 & 1.51 & 7.02 & $\gtorder0.4$ & $18.93\pm0.05$ & 2.66 \\
(99880) 2002 PT91 & 2 & 38 & 3.19 & 2.23 & 5.34 & $\gtorder0.4$ & $19.9\pm0.1$ & 3.02 \\
(100595) 1997 PA2 & 3 & 81 & 3.22 & 2.26 & 5.23 & $\gtorder0.5$ & $18.32\pm0.03$ & 2.94 \\
\enddata
\tablecomments{Columns: asteroids' designations, number of nights, number of images, geocentric ({\it r}) and heliocentric (${\it \Delta}$) distances, phase angle ($\alpha$), lower limit on the lightcurve amplitude, mean magnitude, semi-major axis.}
\label{tab:ResAmpLimit}
\end{deluxetable*}

\begin{deluxetable*}{lcccccccl}
\setcounter{table}{5}
\tablecolumns{9}
\tablewidth{0pt}
\tablecaption{continued}
\tablehead{
\colhead{Designation} &
\colhead{$N_{nights}$} &
\colhead{$N_{im}$} &
\colhead{${\it r}$} &
\colhead{$\Delta$} &
\colhead{$\alpha$} &
\colhead{Amp} &
\colhead{$ \langle R_{PTF} \rangle $} &
\colhead{a} \\
\colhead{}        &
\colhead{}        &
\colhead{}        &
\colhead{[AU]}      &
\colhead{[AU]}      &
\colhead{[deg]} &
\colhead{[mag]}   &
\colhead{[mag]}   &
\colhead{[AU]}
}
\startdata
% Designation    Nnights Nim  r Delta alpha  Amplitude       R_{PTF}    Semi-major axis  
(157971) 2000 GN53 & 3 & 111 & 2.45 & 1.5 & 7.31 & $\gtorder0.2$ & $19.32\pm0.06$ & 2.31 \\
(158514) 2002 EZ116 & 3 & 80 & 2.77 & 1.82 & 6.9 & $\gtorder0.3$ & $19.86\pm0.08$ & 2.62 \\
(158861) 2004 PC8 & 3 & 58 & 2.5 & 1.55 & 7.95 & $\gtorder0.6$ & $18.98\pm0.05$ & 2.35 \\
(158888) 2004 PS71 & 3 & 114 & 2.81 & 1.86 & 7.44 & $\gtorder0.9$ & $19.69\pm0.08$ & 2.39 \\
(159850) 2003 YC35 & 3 & 36 & 3.19 & 2.23 & 5.04 & $\gtorder0.6$ & $18.77\pm0.07$ & 3.05 \\
(160974) 2002 CL38 & 3 & 125 & 2.64 & 1.69 & 7.36 & $\gtorder0.5$ & $18.86\pm0.06$ & 2.55 \\
(161076) 2002 LY40 & 4 & 105 & 3.4 & 2.44 & 4.51 & $\gtorder0.6$ & $19.58\pm0.08$ & 2.8 \\
(161959) 2007 HH48 & 2 & 23 & 2.83 & 1.88 & 7.08 & $\gtorder0.6$ & $19.22\pm0.07$ & 2.44 \\
\enddata
\tablecomments{Columns: asteroids' designations, number of nights, number of images, geocentric ({\it r}) and heliocentric (${\it \Delta}$) distances, phase angle ($\alpha$), lower limit on the lightcurve amplitude, mean magnitude, semi-major axis.}
\label{tab:ResAmpLimit}
\end{deluxetable*}

\begin{deluxetable*}{lcccccccl}
\setcounter{table}{5}
\tablecolumns{9}
\tablewidth{0pt}
\tablecaption{continued}
\tablehead{
\colhead{Designation} &
\colhead{$N_{nights}$} &
\colhead{$N_{im}$} &
\colhead{${\it r}$} &
\colhead{$\Delta$} &
\colhead{$\alpha$} &
\colhead{Amp} &
\colhead{$ \langle R_{PTF} \rangle $} &
\colhead{a} \\
\colhead{}        &
\colhead{}        &
\colhead{}        &
\colhead{[AU]}      &
\colhead{[AU]}      &
\colhead{[deg]} &
\colhead{[mag]}   &
\colhead{[mag]}   &
\colhead{[AU]}
}
\startdata
% Designation    Nnights Nim  r Delta alpha  Amplitude       R_{PTF}    Semi-major axis  
(230887) 2004 RW345 & 3 & 24 & 2.4 & 1.45 & 8.55 & $\gtorder0.3$ & $18.37\pm0.03$ & 2.69 \\
(230911) 2004 TR201 & 2 & 23 & 2.45 & 1.49 & 7.84 & $\gtorder0.6$ & $19.46\pm0.08$ & 2.68 \\
(231151) 2005 UR41 & 3 & 44 & 2.27 & 1.32 & 8.75 & $\gtorder0.5$ & $17.99\pm0.04$ & 2.34 \\
(231297) 2006 BT141 & 2 & 20 & 2.82 & 1.87 & 6.37 & $\gtorder0.4$ & $19.48\pm0.08$ & 2.66 \\
(231463) 2007 PO & 2 & 23 & 3.01 & 2.06 & 5.8 & $\gtorder0.3$ & $20.4\pm0.1$ & 3.14 \\
(231532) 2008 SK126 & 1 & 12 & 2.72 & 1.76 & 5.83 & $\gtorder0.6$ & $20.6\pm0.1$ & 2.94 \\
(231555) 2008 TT2 & 3 & 30 & 3.02 & 2.06 & 5.51 & $\gtorder0.8$ & $20.7\pm0.1$ & 2.93 \\
(237344) 2010 AY23 & 3 & 46 & 2.03 & 1.07 & 9.48 & $\gtorder0.8$ & $19.52\pm0.08$ & 2.23 \\
\enddata
\tablecomments{Columns: asteroids' designations, number of nights, number of images, geocentric ({\it r}) and heliocentric (${\it \Delta}$) distances, phase angle ($\alpha$), lower limit on the lightcurve amplitude, mean magnitude, semi-major axis.}
\label{tab:ResAmpLimit}
\end{deluxetable*}

\begin{deluxetable*}{lcccccccl}
\setcounter{table}{5}
\tablecolumns{9}
\tablewidth{0pt}
\tablecaption{continued}
\tablehead{
\colhead{Designation} &
\colhead{$N_{nights}$} &
\colhead{$N_{im}$} &
\colhead{${\it r}$} &
\colhead{$\Delta$} &
\colhead{$\alpha$} &
\colhead{Amp} &
\colhead{$ \langle R_{PTF} \rangle $} &
\colhead{a} \\
\colhead{}        &
\colhead{}        &
\colhead{}        &
\colhead{[AU]}      &
\colhead{[AU]}      &
\colhead{[deg]} &
\colhead{[mag]}   &
\colhead{[mag]}   &
\colhead{[AU]}
}
\startdata
% Designation    Nnights Nim  r Delta alpha  Amplitude       R_{PTF}    Semi-major axis  
2006 JM42 & 2 & 22 & 3.41 & 2.47 & 5.62 & $\gtorder0.4$ & $20.5\pm0.1$ & 3.05 \\
2006 KP31 & 3 & 37 & 2.75 & 1.8 & 7.19 & $\gtorder0.4$ & $20.4\pm0.1$ & 2.94 \\
2007 EE153 & 3 & 63 & 2.01 & 1.05 & 9.72 & $\gtorder0.7$ & $19.75\pm0.07$ & 2.18 \\
2007 EA198 & 1 & 8 & 2.4 & 1.44 & 7.74 & $\gtorder0.3$ & $20.7\pm0.2$ & 2.27 \\
2007 HE51 & 3 & 22 & 2.33 & 1.38 & 8.39 & $\gtorder0.5$ & $20.2\pm0.1$ & 2.39 \\
2007 JY18 & 2 & 25 & 2.21 & 1.26 & 8.62 & $\gtorder0.5$ & $20.6\pm0.1$ & 2.36 \\
2007 JY22 & 1 & 16 & 2.95 & 2 & 6.53 & $\gtorder0.7$ & $20.4\pm0.1$ & 2.59 \\
2007 JM24 & 2 & 51 & 2.69 & 1.74 & 7.17 & $\gtorder0.4$ & $19.88\pm0.1$ & 2.55 \\
\enddata
\tablecomments{Columns: asteroids' designations, number of nights, number of images, geocentric ({\it r}) and heliocentric (${\it \Delta}$) distances, phase angle ($\alpha$), lower limit on the lightcurve amplitude, mean magnitude, semi-major axis.}
\label{tab:ResAmpLimit}
\end{deluxetable*}

\begin{deluxetable*}{lcccccccl}
\setcounter{table}{5}
\tablecolumns{9}
\tablewidth{0pt}
\tablecaption{continued}
\tablehead{
\colhead{Designation} &
\colhead{$N_{nights}$} &
\colhead{$N_{im}$} &
\colhead{${\it r}$} &
\colhead{$\Delta$} &
\colhead{$\alpha$} &
\colhead{Amp} &
\colhead{$ \langle R_{PTF} \rangle $} &
\colhead{a} \\
\colhead{}        &
\colhead{}        &
\colhead{}        &
\colhead{[AU]}      &
\colhead{[AU]}      &
\colhead{[deg]} &
\colhead{[mag]}   &
\colhead{[mag]}   &
\colhead{[AU]}
}
\startdata
% Designation    Nnights Nim  r Delta alpha  Amplitude       R_{PTF}    Semi-major axis  
P000JD & 3 & 63 & \nodata & \nodata & \nodata & $\gtorder1.7$ & $19.59\pm0.07$ & \nodata \\
P000JF & 2 & 54 & \nodata & \nodata & \nodata & $\gtorder1.3$ & $20.3\pm0.1$ & \nodata \\
P000JQ & 3 & 60 & \nodata & \nodata & \nodata & $\gtorder0.3$ & $19.79\pm0.09$ & \nodata \\
P000JX & 2 & 59 & \nodata & \nodata & \nodata & $\gtorder0.6$ & $19.96\pm0.1$ & \nodata \\
P000K1 & 1 & 23 & \nodata & \nodata & \nodata & $\gtorder0.8$ & $19.13\pm0.06$ & \nodata \\
P000K2 & 1 & 26 & \nodata & \nodata & \nodata & $\gtorder0.4$ & $18.99\pm0.04$ & \nodata \\
P000L7 & 1 & 15 & \nodata & \nodata & \nodata & $\gtorder2$ & $20.1\pm0.1$ & \nodata \\
P000LL & 2 & 29 & \nodata & \nodata & \nodata & $\gtorder0.4$ & $20.2\pm0.1$ & \nodata \\
\enddata
\tablecomments{Columns: asteroids' designations, number of nights, number of images, geocentric ({\it r}) and heliocentric (${\it \Delta}$) distances, phase angle ($\alpha$), lower limit on the lightcurve amplitude, mean magnitude, semi-major axis.}
\label{tab:ResAmpLimit}
\end{deluxetable*}

\begin{deluxetable*}{lcccccccl}
\setcounter{table}{5}
\tablecolumns{9}
\tablewidth{0pt}
\tablecaption{continued}
\tablehead{
\colhead{Designation} &
\colhead{$N_{nights}$} &
\colhead{$N_{im}$} &
\colhead{${\it r}$} &
\colhead{$\Delta$} &
\colhead{$\alpha$} &
\colhead{Amp} &
\colhead{$ \langle R_{PTF} \rangle $} &
\colhead{a} \\
\colhead{}        &
\colhead{}        &
\colhead{}        &
\colhead{[AU]}      &
\colhead{[AU]}      &
\colhead{[deg]} &
\colhead{[mag]}   &
\colhead{[mag]}   &
\colhead{[AU]}
}
\startdata
% Designation    Nnights Nim  r Delta alpha  Amplitude       R_{PTF}    Semi-major axis  
P001P1 & 1 & 7 & \nodata & \nodata & \nodata & $\gtorder0.3$ & $20.4\pm0.1$ & \nodata \\
P001T1 & 3 & 73 & \nodata & \nodata & \nodata & $\gtorder0.0$ & $19.95\pm0.09$ & \nodata \\
P001TQ & 3 & 79 & \nodata & \nodata & \nodata & $\gtorder0.3$ & $20.4\pm0.1$ & \nodata \\
P001U2 & 2 & 69 & \nodata & \nodata & \nodata & $\gtorder0.4$ & $19.26\pm0.05$ & \nodata \\
P001U4 & 2 & 30 & \nodata & \nodata & \nodata & $\gtorder0.4$ & $19.9\pm0.1$ & \nodata \\
P001VG & 2 & 28 & \nodata & \nodata & \nodata & $\gtorder0.9$ & $20.2\pm0.1$ & \nodata \\
P001VX & 2 & 8 & \nodata & \nodata & \nodata & $\gtorder1.1$ & $20.6\pm0.2$ & \nodata \\
P001WD & 3 & 41 & \nodata & \nodata & \nodata & $\gtorder1.6$ & $20.2\pm0.1$ & \nodata \\
\enddata
\tablecomments{Columns: asteroids' designations, number of nights, number of images, geocentric ({\it r}) and heliocentric (${\it \Delta}$) distances, phase angle ($\alpha$), lower limit on the lightcurve amplitude, mean magnitude, semi-major axis.}
\label{tab:ResAmpLimit}
\end{deluxetable*}

\title{Asteroid rotation periods from PTF survey: Supplementary Material}

\begin{widetext}  
\noindent Table 6 can be found in its entirety here:

\noindent $http://wise-obs.tau.ac.il/\sim david/PTF/Data/Polishook\_EtAl\_2012\_MNRAS\_PTF\_Asteroids\_Table6$

\noindent Columns: asteroids’ designations, PTF’s field and CCD, JD [days], RA [degrees], Dec [degrees], calibrated magnitude [mag], air mass, image’s background count and FWHM [arcsec]. The last six parameters are given only for objects with known orbital parameters: geocentric (r) and heliocentric ($\Delta$) distances [AU], phase angle ($\alpha$) [degrees], semi-major axis [AU], eccentricity and inclination [degrees].
\end{widetext}